\pdfoutput=1
\documentclass[3p, onecolumn,authoryear, final]{elsarticle}

\usepackage{url}
\usepackage{hyperref}
\usepackage{amsthm}
\usepackage{caption}
\usepackage{verbatim}
\usepackage{algorithm}
\usepackage{algorithmic}
\usepackage{enumitem}
\usepackage{color}
\usepackage[export]{adjustbox}

\graphicspath{{./color.png/}}
\DeclareGraphicsExtensions{.png}

\usepackage[caption=false,font=footnotesize,labelfont=sf,textfont=sf]{subfig}

\newtheorem{definition}{Definition}
\newtheorem{proposition}{Proposition}
\newcommand{\s}{0.35}
\newcommand{\sw}{0.3}

\journal{}

\biboptions{authoryear}

\newcommand{\algname}{\textit}

\begin{document}

\begin{frontmatter}

\title{Revisiting 2-3 Red-Black Trees with a Pedagogically Sound yet Efficient Deletion Algorithm: The Parity-Seeking Delete Algorithm
}

\author[address]{Kamaledin Ghiasi-Shirazi\corref{mycorrespondingauthor}}
\cortext[mycorrespondingauthor]{Corresponding author}
\ead{k.ghiasi@um.ac.ir}

\author[address]{Taraneh Ghandi}
\ead{taraneh.ghandi@mail.um.ac.ir}

\author[address]{Ali Taghizadeh}
\ead{ali.taghizadeh@mail.um.ac.ir}

\author[address]{Ali Rahimi-Baigi}
\ead{alirahimibaigi@mail.um.ac.ir}

\address[address]{Computer Engineering Department, Ferdowsi University of Mashhad, Mashhad, Iran}

\begin{abstract}
Red-black (RB) trees are one of the most efficient variants of balanced binary search trees. However, they have always been blamed for being too complicated, hard to explain, and not suitable for pedagogical purposes. 
In the pioneering work of \citet{guibas1978dichromatic}, both 2-3 and 2-3-4 variants of RB trees had been considered, but further study of the former had been abandoned due to the higher number of rotations in the \algname{insert} algorithm.
\citet{sedgewick2008left} proposed a variant of 2-3 RB trees, viz. left-leaning red-black (LLRB) trees, in which red links are restricted to left children and proposed concise recursive \algname{insert} and \algname{delete} algorithms. However, the top-down deletion algorithm of LLRB is still very complicated and highly inefficient.
In this paper, we reconsider 2-3 red-black trees in which both children of a node cannot be red. We propose a parity-seeking \algname{delete} algorithm with the basic idea of making the deficient subtree on a par with its sibling: either by fixing the deficient subtree or by turning the sibling deficient as well, ascending deficiency to the parent node. 
Interestingly, the proposed parity-seeking \algname{delete} algorithm works for 2-3-4 RB trees as well.
Our experiments show that 2-3 RB trees are almost as efficient as RB trees and twice faster than LLRB trees.
Besides, RB trees with the proposed parity-seeking \algname{delete} algorithm have the same number of rotations and almost identical running time as the classical \algname{delete} algorithm. 
While being extremely efficient, the proposed parity-seeking \algname{delete} algorithm is easily understandable and suitable for pedagogical purposes.
\end{abstract}


\end{frontmatter}


\section{Introduction}
\citet{bayer1970organization, bayer1972organization} invented B-trees which are balanced-tree data structures appropriate for the organization and maintenance of large ordered indices, especially on disks.
Since each node of a B-tree should allocate room for a predetermined maximum number of keys, B-trees are memory-inefficient.
By chaining the keys of a B-tree node by horizontal links (termed $\rho$-arcs in their terminology), \citet{bayer1971binary} introduced a binary tree representation of B-trees which avoided their storage overhead.
\citet{bayer1972symmetric}  introduced symmetric binary trees, which were binary tree representations of 2-3-4 trees and allowed the keys within a B-tree to be either linked by left or right pointers. 
Symmetric binary trees were named red-black (RB) trees thereafter when \citet{guibas1978dichromatic} proposed a dichromatic framework for balanced trees.
\citet{guibas1978dichromatic} investigated \algname{insert} algorithms for both 2-3 and 2-3-4 variants of RB trees, but abandoned further study of 2-3 RB trees due to their slightly higher number of rotations.
Since then, many improvements to RB trees have been proposed.
Some authors \citep{andersson1990binary, roura2013fibonacci} tried to decrease the maximum height of RB trees, which is $2 \log(n)$ in the worst case. 
Others tried to uncouple updating from rebalancing, allowing a greater degree of concurrency and postponed processing \citep{boyar1994efficient, park2001parallel, larsen2002relaxed, besa2013concurrent, howard2014relativistic}.

While being extremely useful in applications, RB trees have always been criticized for being baffling and inappropriate for pedagogical purposes.
To simplify RB trees, \citet{andersson1993balanced} proposed right-leaning red-black trees in which only the right nodes could be red. 
However, their method works differently and requires explicit storage of level numbers in the nodes.
In another attempt to simplify RB trees, \citet{okasaki1999red} proposed an algorithm for insertion into RB trees using functional programming in Haskell.
By temporarily introducing a third "double-black" color, \citet{germane2014deletion} proposed a functional \algname{delete} algorithm for RB trees.
Attempting to simplify RB trees for pedagogical purposes, \citet{sedgewick2008left} proposed left-leaning red-black (LLRB) trees ---a variant of 2-3 RB trees--- in which red links are only permitted on the left.
Although the \algname{insert} algorithm of LLRB trees is simple, we found the \algname{delete} algorithm very unintuitive and difficult to understand \footnote{We have visualized the steps taken by each variant of RB trees on randomly inserting and deleting 30 numbers. This visualization is available at \url{https://profsite.um.ac.ir/~k.ghiasi/publications/PS-RBT/index.html}.}. 
In fact, the real problem with classical RB trees is the \algname{delete} algorithm due to its unclear rationale \citep{germane2014deletion, sen2016deletion}.

In this paper, we reconsider 2-3 RB trees, in which the children of a node cannot both be red, and propose a novel insertion algorithm and an intuitive parity-seeking \algname{delete} algorithm that are highly suitable for educational purposes. We then show that the proposed parity-seeking \algname{delete} algorithm can also be used in ordinary 2-3-4 RB trees, yielding a pedagogically sound \algname{delete} algorithm for RB trees. Besides, we show that the slightly higher number of rotations in the \algname{insert} algorithm has completely negligible effect on the performance of 2-3 RB trees. Our experiments on 2-3 and 2-3-4 RB trees show that the proposed parity-seeking \algname{delete} algorithm is extremely efficient.

The rest of the paper proceeds as follows: 
In Section~\ref{sec:rbt}, we review the classical algorithm of RB trees as  explained in \citet[Chapter~13]{cormen2009introduction}.
In Section~\ref{sec:llrb}, we review LLRB trees \citep{sedgewick2008left} and argue that, despite having fewer lines of code, the deletion algorithm is very inefficient and unintuitive.
In Section~\ref{sec:2-3 RB}, we reconsider 2-3 RB trees and propose an insertion algorithm along with a novel parity-seeking \algname{delete} algorithm that is much simpler than the \algname{delete} algorithm of classical RB trees.
In Section~\ref{sec:clarification}, we 
show that the proposed parity-seeking \algname{delete} algorithm works, without modification, for RB trees as well.
In Section~\ref{sec:experiments}, we experimentally evaluate the performance of the standard RB trees, as described by \citet[Chapter~13]{cormen2009introduction}, LLRB, and the proposed 2-3 and 2-3-4 RB trees.
We conclude the paper in Section~\ref{sec:conclusions}.

\section{Red-Black (RB) trees}\label{sec:rbt}

\begin{definition}[RB trees]
\label{def:rbt}
An RB tree is a binary search tree with one additional attribute in each node: its color, which can be either red or black. RB trees have the following properties:
\begin{enumerate}
\item The root node is black,
\item If a node is red, then its parent is black,
\item Each path from the root to a leaf contains the same number of black nodes (called the black height of the tree).
\end{enumerate}
\end{definition}
We replace the null pointers of leaf and degree-1 nodes by pointers to some imaginary black nodes called external nodes. More precisely, we represent all external nodes by a common black sentinel node \citep[Chapter~13]{cormen2009introduction}, called the \textit{nilSentinel} node. 
In addition, as in \citep[Chapter~13]{cormen2009introduction}, the left child of the \textit{nilSentinel} node is set to the root node and its right child is set to itself. 
For an empty tree, the left child of the \textit{nilSentinel} node is also set to itself.
Sometimes it is useful to refer to the color of a link. 
The color of the link from a parent to a child node, is the color of the child node.
In all illustrations of this paper, we depict black nodes and links by solid lines, the red nodes and links by solid double lines, and those that could be either red or black by dotted lines.

\subsection{Relation between RB trees and B-trees of order 4 (2-3-4 trees)}
\begin{figure}[h]%
\centering
\subfloat[]{\includegraphics[scale=\s, valign=c]{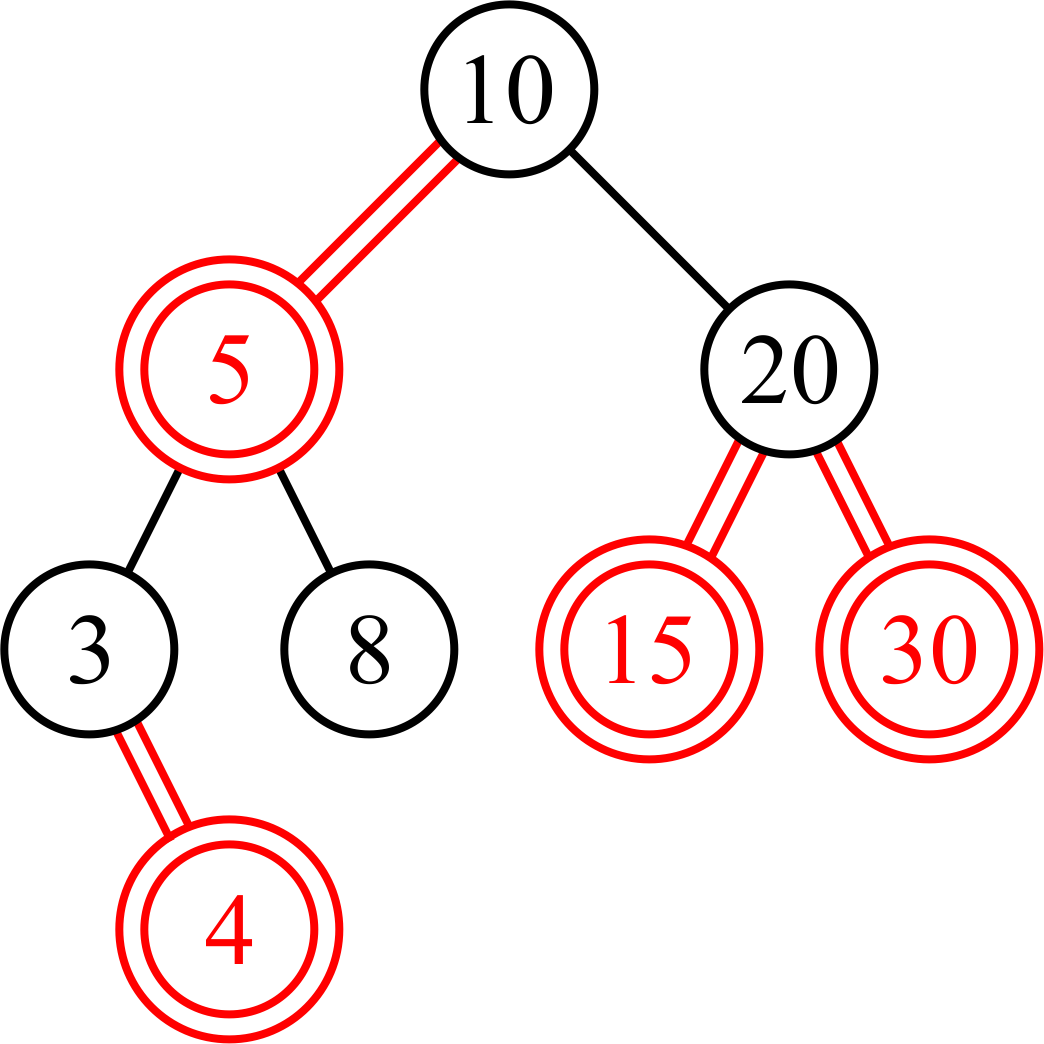}}
\qquad
\subfloat[]{\includegraphics[scale=\s, valign=c]{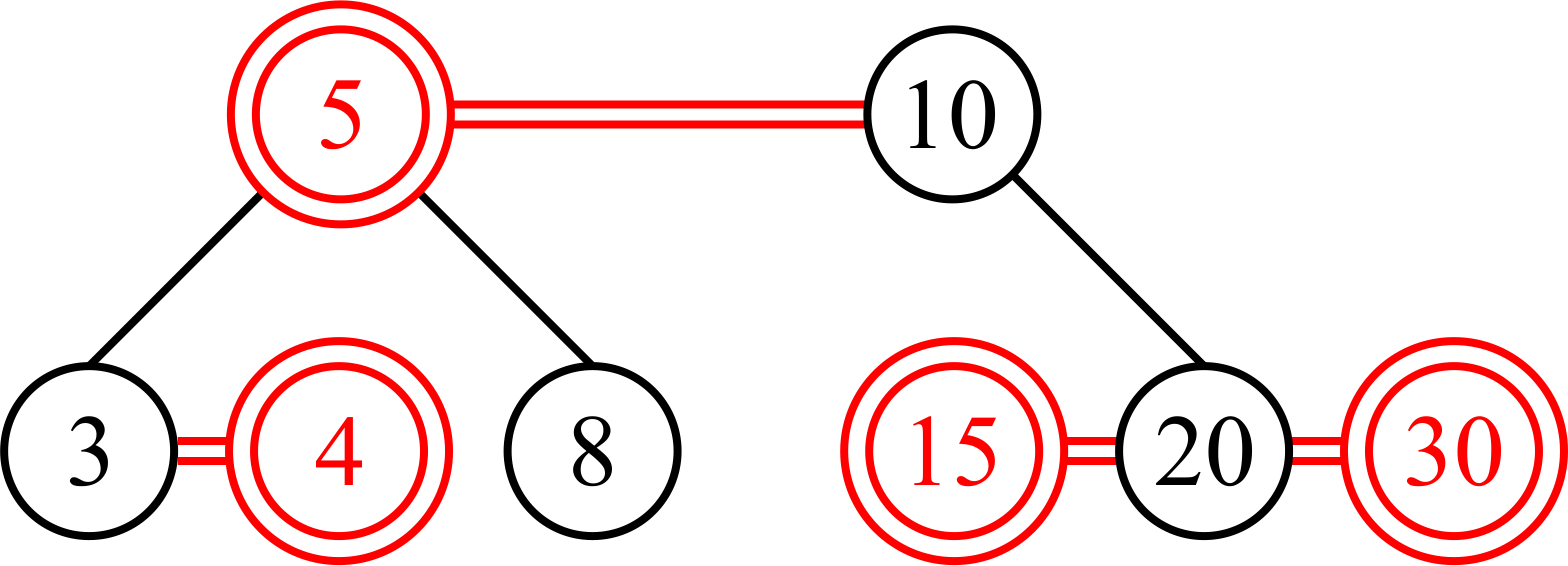}}\qquad
\subfloat[]{\includegraphics[scale=\s, valign=c]{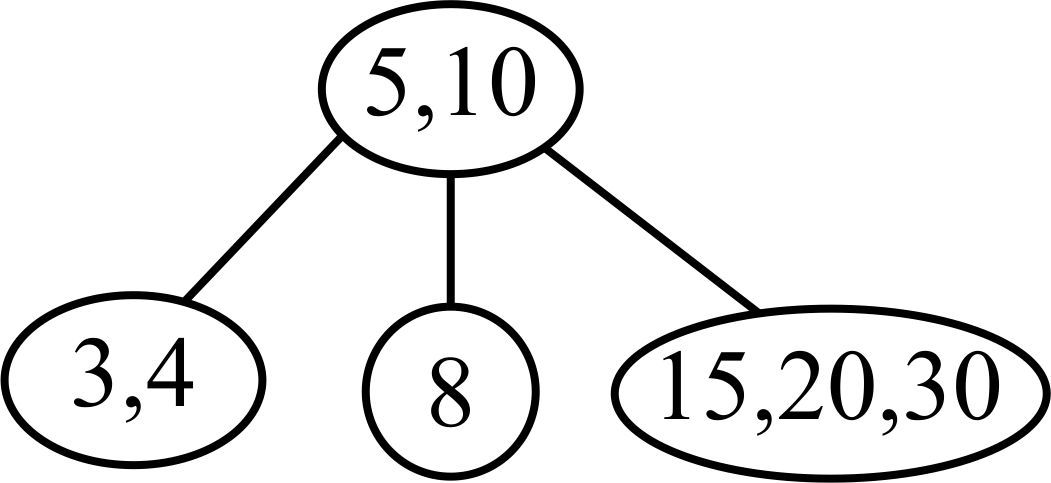}}
\caption{(a) An RB tree, (b) its representation with horizontal red links, and (c) its equivalent 2-3-4 tree.}
\label{fig:rbt-234-equiv}
\end{figure}

Considering an RB tree, if we draw the red links horizontally and the black links vertically, then a representation is obtained in which, due to the 3rd property in Definition~\ref{def:rbt}, all leaves are drawn at the same level. Furthermore, if we place the horizontally connected nodes in one compound node, then the 2-3-4 tree equivalent of the very RB tree is obtained.
Figure~\ref{fig:rbt-234-equiv} shows an RB tree along with its other equivalent representations.

\subsection{Basic operations in RB trees}
\begin{figure}[h]%
\centering
\includegraphics[scale=\s,valign=c]{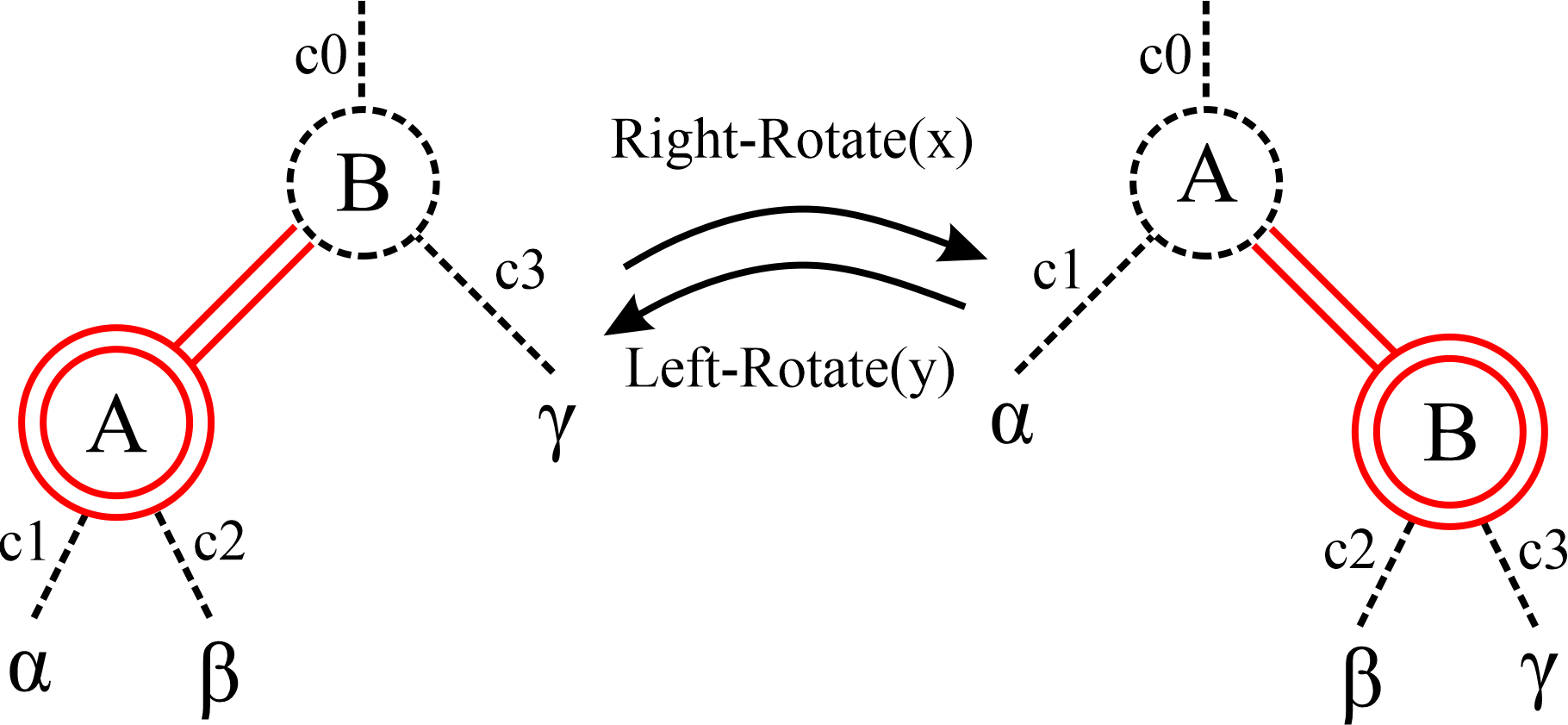}
\caption{Left rotation and right rotation. Here $\alpha,\beta$, and $\gamma$ represent subtrees. Some of the nodes and links are dotted to show that their color is not known. The color of links are symbolically shown by letters $c0,...,c3$. }
\label{fig:rotations}
\end{figure}

After inserting/deleting a node into/from an RB tree, the properties of Definition~\ref{def:rbt} might become violated.
While modifying the tree in order to comply with Definition~\ref{def:rbt}, it is important that the order of the nodes in the inorder traversal of the tree does not change, so that the resulting tree remains a valid binary search tree.
In this section, we introduce the basic operations that preserve the properties of binary search trees. These operations are left rotation and right rotation, which are shown in Figure~\ref{fig:rotations}.
Furthermore, changing the color of nodes is another operation that preserves the properties of binary search trees.
To understand the color of nodes after rotation, it is easier to assume that the links are rotated and infer the color of nodes from the color of their links to their parents.

\subsection{Insertion algorithm of RB trees}
\begin{figure}[h]%
\centering
\subfloat[]{\includegraphics[scale=\s, valign=c]{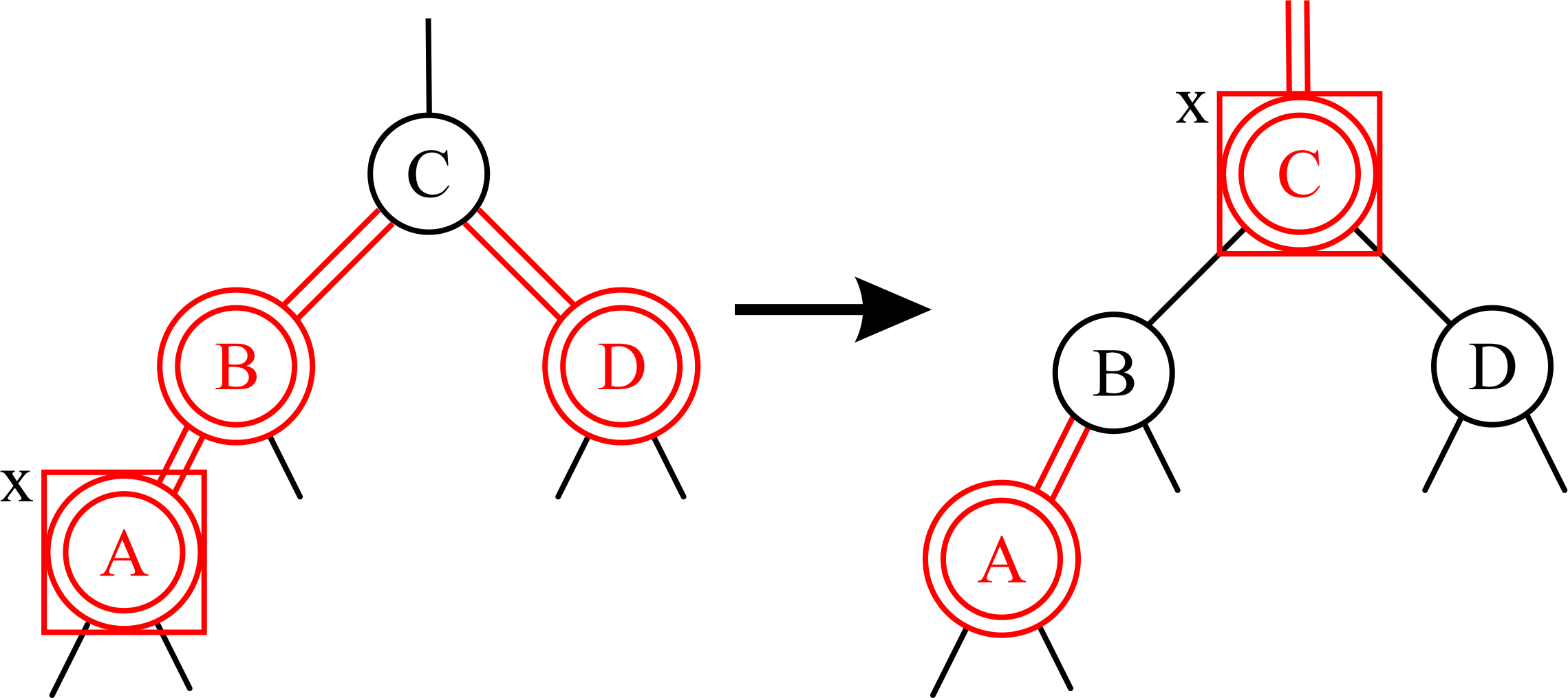}
\label{fig:rbt-insert-case-1}}\qquad\qquad
\subfloat[]{\includegraphics[scale=\s, valign=c]{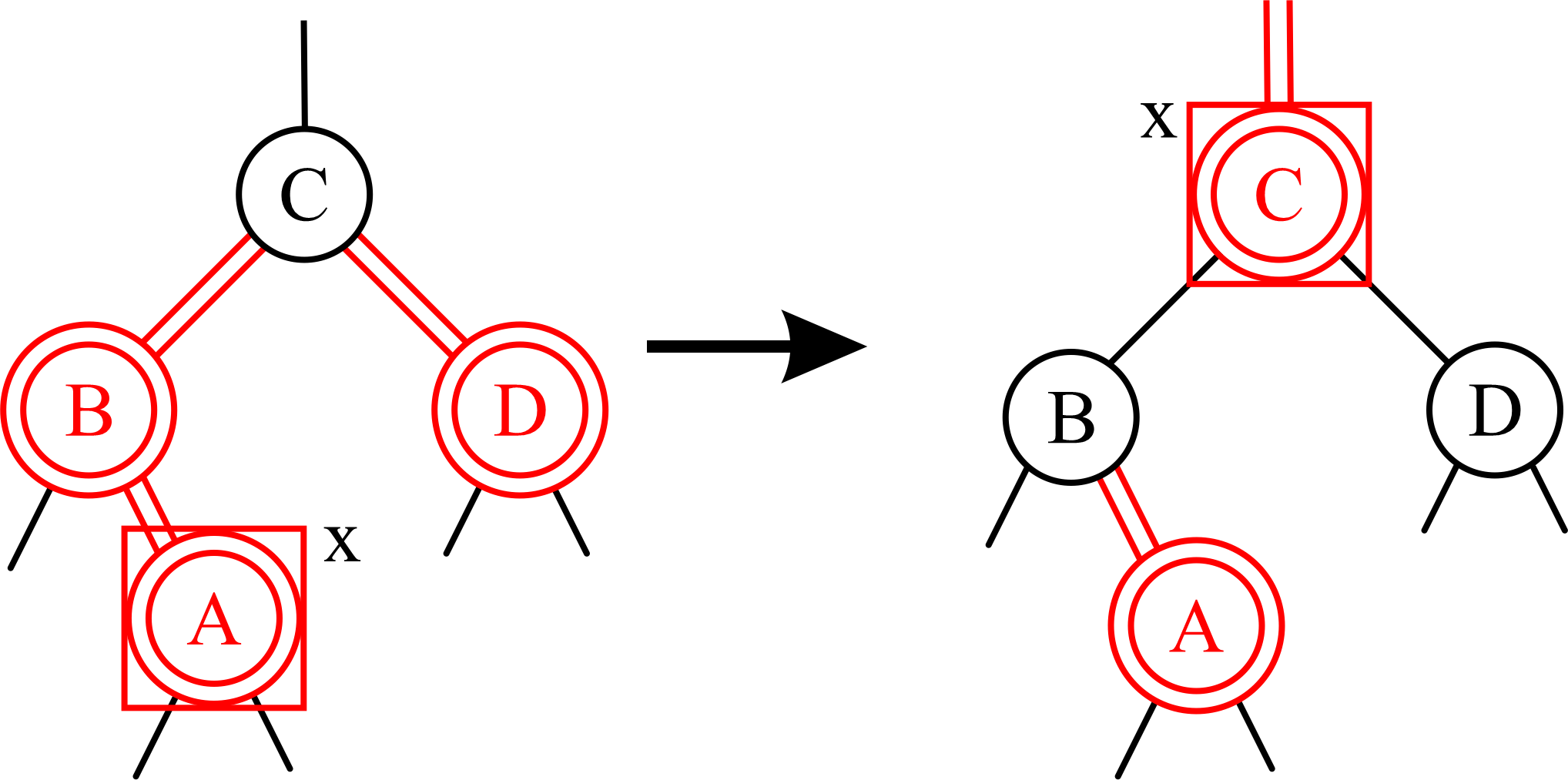}
\label{fig:rbt-insert-case-2}}\qquad\qquad
\subfloat[]{\includegraphics[scale=\s, valign=c]{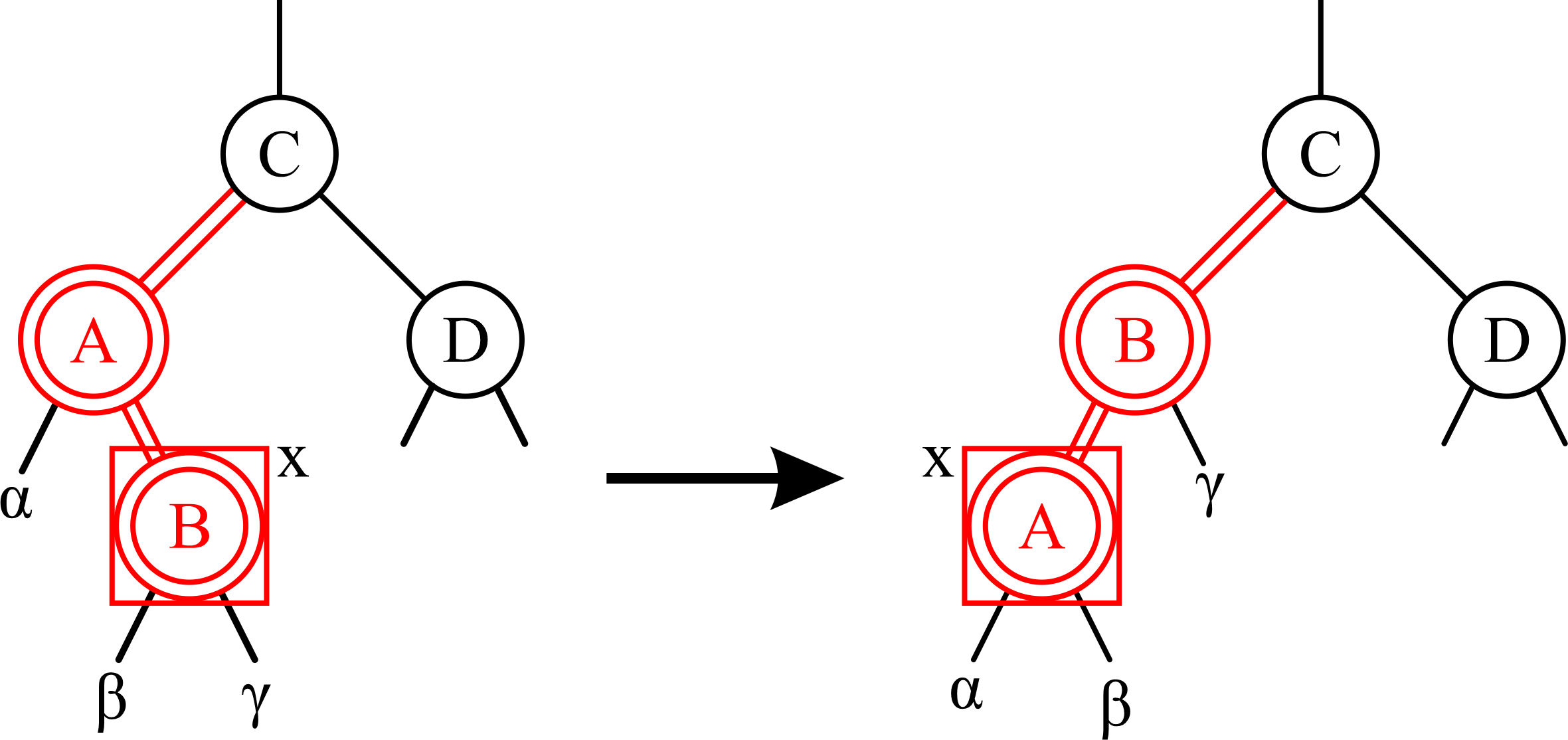}
\label{fig:rbt-insert-case-3}}
\subfloat[]{\includegraphics[scale=\s, valign=c]{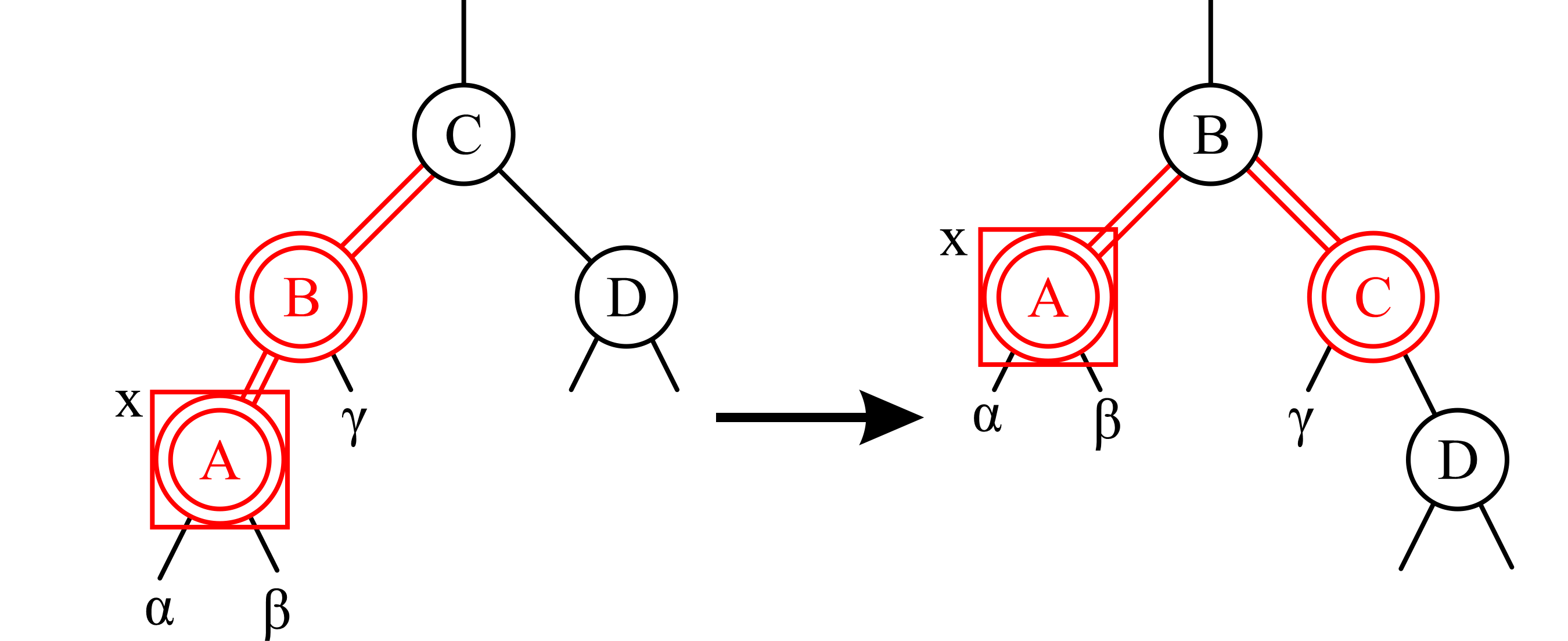}
\label{fig:rbt-insert-case-4}}
\caption{Rules for fixing RB trees after inserting a new node. The rules are applied recursively. Only half of the rules, for the case that the parent of the inserted node is a left child, are shown. The other $4$ rules are obtained by symmetry. Subtrees are shown by Greek letters and red nodes and links are shown by double lines. The node under consideration is denoted by the letter $x$ and a square. Note that, the implementation of RB trees  considers only three cases, and rules (a) and (b) are handled jointly. 
}
\label{fig:rbt-insert} 
\end{figure}

The \algname{insert} algorithm of RB trees works in two passes: top-down pass and bottom-up pass. During the top-down pass, the new data is inserted according to the rules of binary search trees in a new red node.
Then, during the bottom-up pass, the tree is fixed with appropriate fix-up operations to ensure that all properties of Definition~\ref{def:rbt} are held.
The 1st property of Definition~\ref{def:rbt} is ensured by setting the color of the root node to black at the end of the algorithm.
The 3rd property of Definition~\ref{def:rbt} would not be violated as the newly inserted node is colored red and all fix-up operations keep this property. 
The only potential problem is the violation of the 2nd property of Definition~\ref{def:rbt}, i.e. the color of a node and its parent both being red.
We use $x$ to signify the sole red node that might have a red parent. 
Initially, $x$ is the newly inserted node.
Assuming that the parent of $x$ is a left child, the tree is fixed using the following rules:
\begin{enumerate}
\item If the sibling of the parent node is red, then the parent node and its sibling are turned black and the grandparent node is turned red. Checking for two consecutive red-nodes is continued from the grandparent node (Figure~\ref{fig:rbt-insert-case-1} and Figure~\ref{fig:rbt-insert-case-2}).
\item If the sibling of the parent node is black, and the current node is a right child, then a left rotation is performed on the parent node (Figure~\ref{fig:rbt-insert-case-3}). The situation becomes ready for applying the next rule.
\item If the sibling of the parent node is black, and the current node is a left child, then a right rotation is performed on the grandparent node (Figure~\ref{fig:rbt-insert-case-4}).
\end{enumerate}
The rules for the case that the parent node is a right child, are obtained by exchanging "left" and "right" in the above statements. The bottom-up pass terminates as soon as the parent of $x$ is not red.

\subsection{Deletion algorithm of RB trees}
\begin{figure}[h]%
\centering
\subfloat[]{\includegraphics[scale=\s, valign=c]{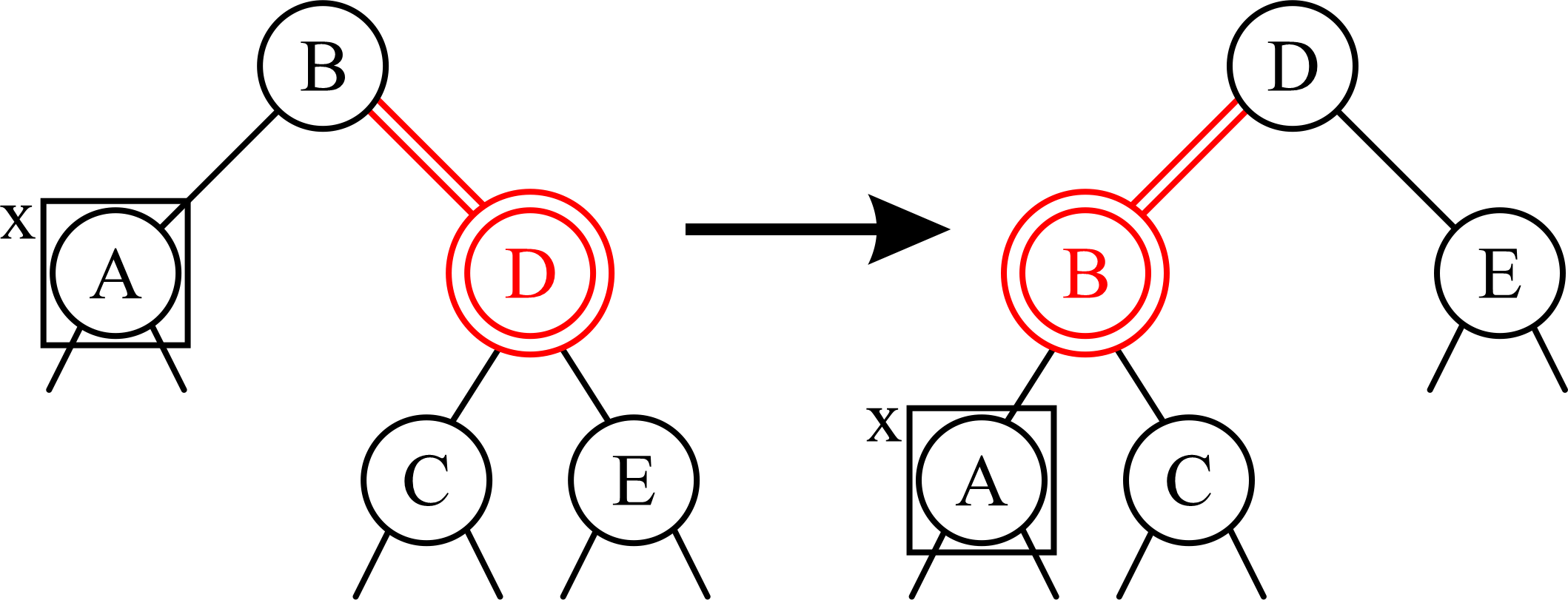}
\label{fig:rbt-delete-case-3-1}
}
\hspace{2cm}
\subfloat[]{\includegraphics[scale=\s, valign=c]{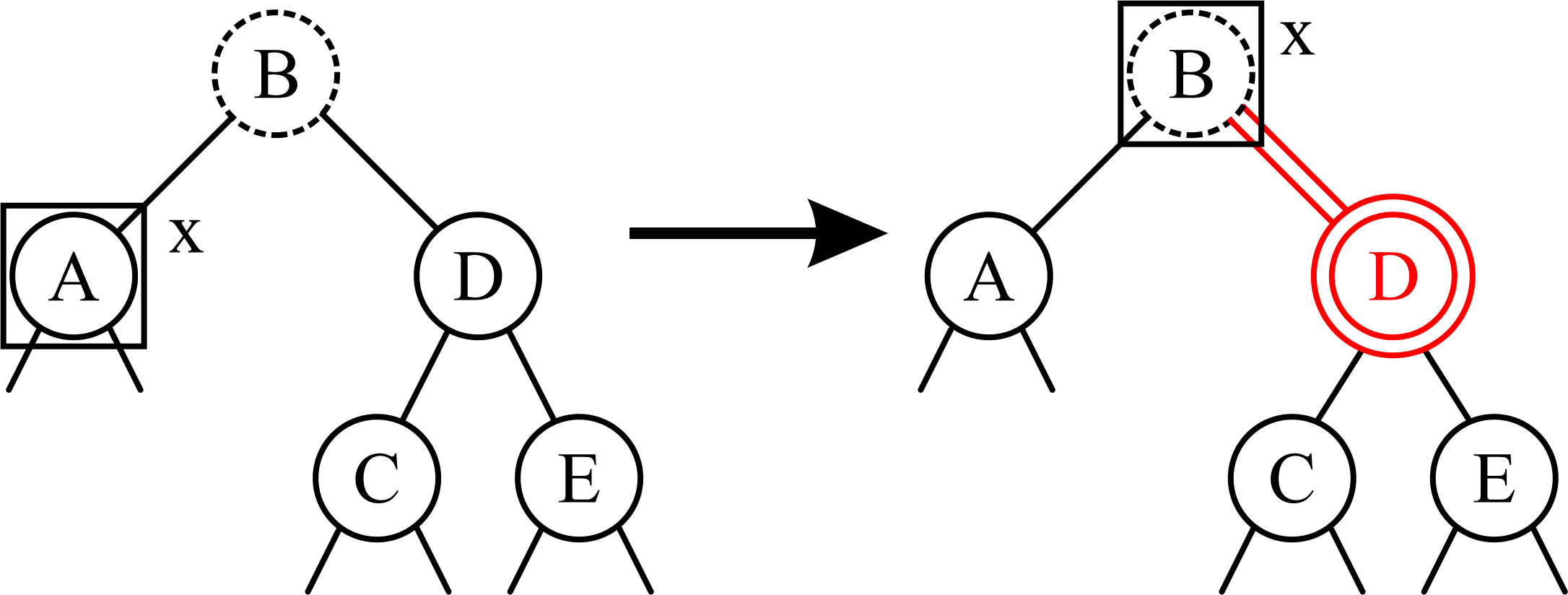}
\label{fig:rbt-delete-case-3-2}
}\\
\vspace{0.4cm}
\subfloat[]{
\includegraphics[scale=\s, valign=c]{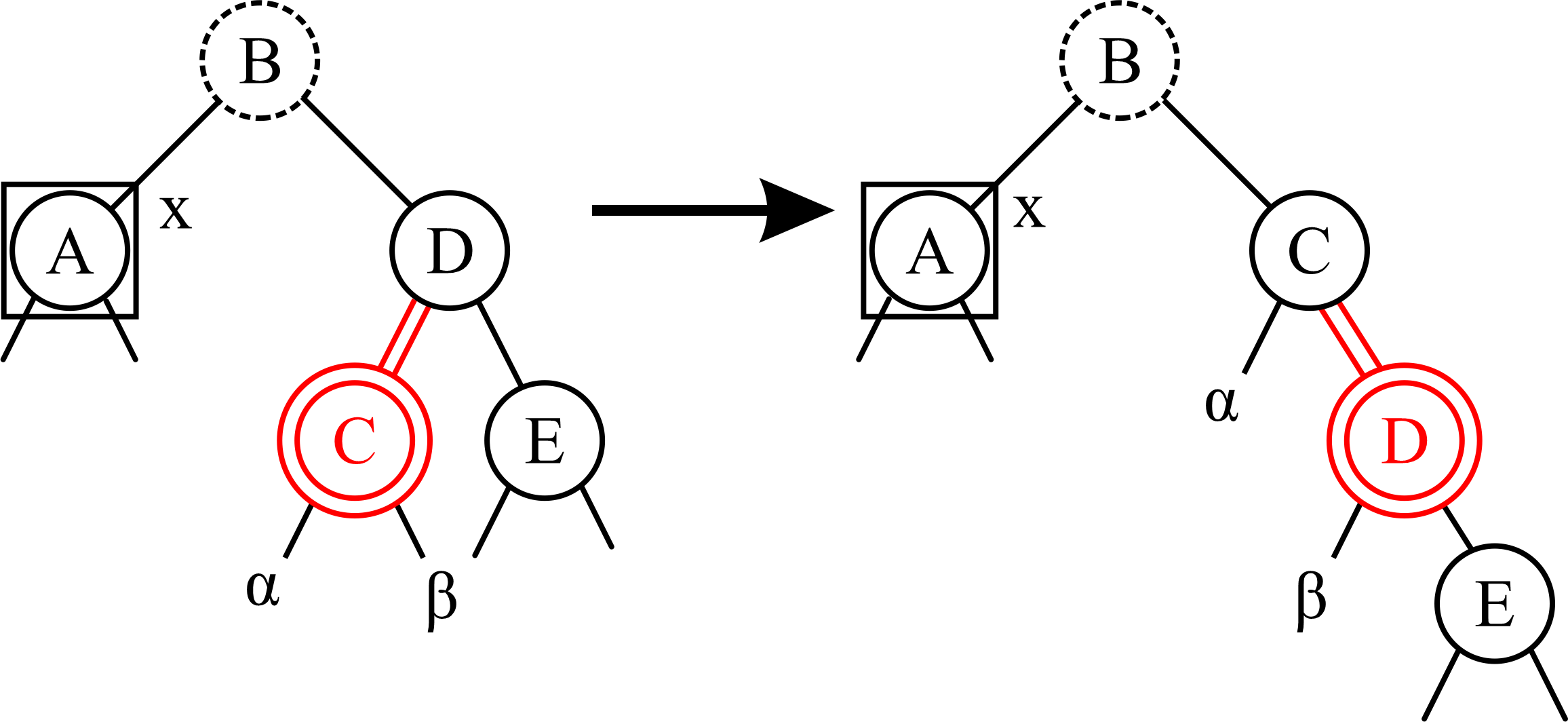}
\label{fig:rbt-delete-case-3-3}
}\hspace{2cm}
\subfloat[]{
\includegraphics[scale=\s, valign=c]{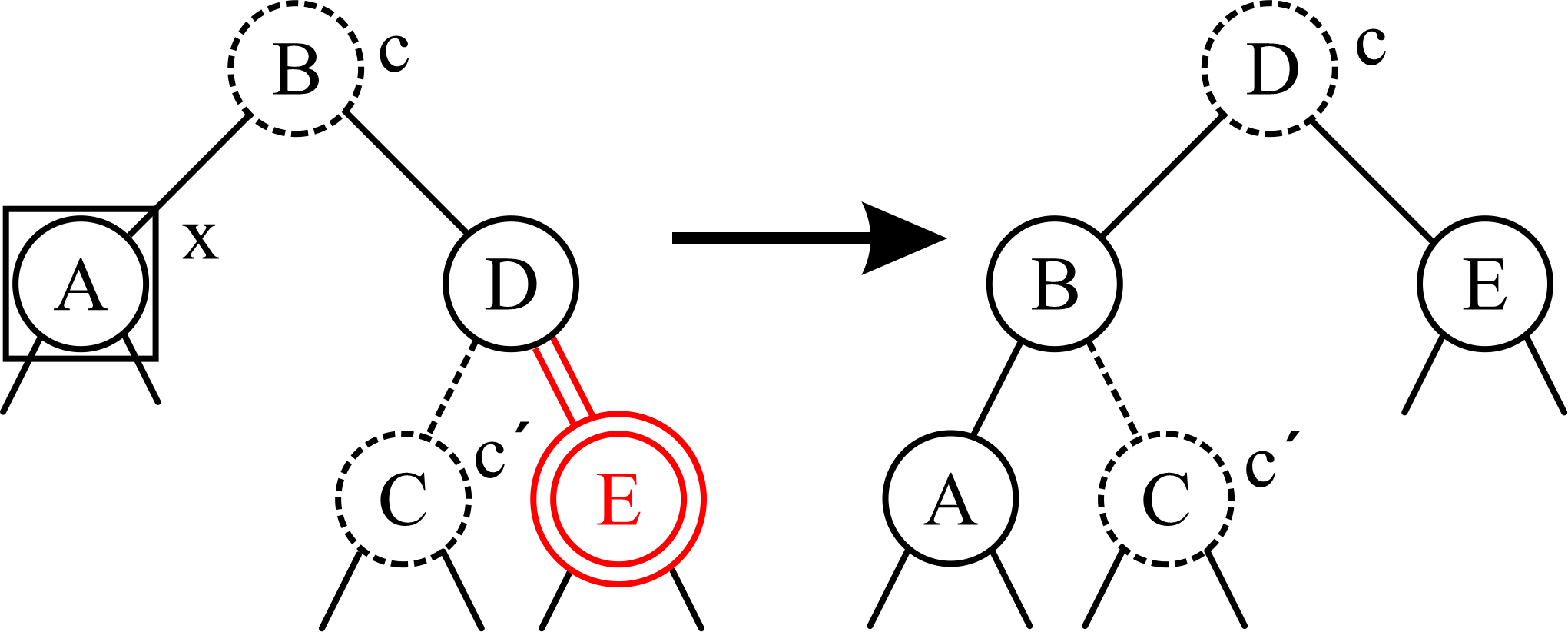}
\label{fig:rbt-delete-case-3-4}
}
\vspace{0.4cm}
\subfloat[]{
\includegraphics[scale=\s, valign=c]{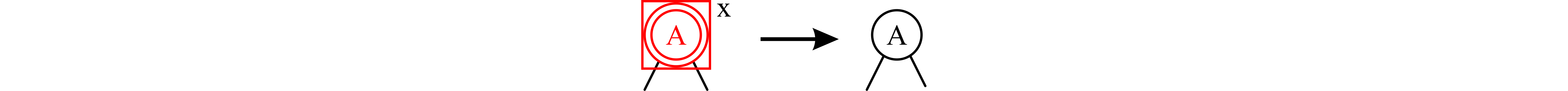}
\label{fig:rbt-delete-case-3-5}
}
\caption{Rules for fixing RB trees after deleting a black leaf node. The rules are applied recursively. The root of the deficient subtree, whose black height is one less than its sibling, is shown with a square and is called $x$. Each rule, except (e), has a dual rule which is obtained by symmetry and is not shown. 
Rule (a) prepares a black sibling for $x$, preparing the situation for a subsequent (c) or (d) step. Rule (b) applies when the sibling and both of its children are black. It elevates deficiency to the parent node. 
Rule (c) applies when the sibling and its right child are black, but the left child of the sibling  is red. It makes the right child of the sibling red and prepares the situation for the next rule. Rule (d) applies when the sibling is black and its right child is red. After applying this rule, the deficiency is removed altogether and the algorithm terminates.
Rule (e) applies when the root of the deficient subtree becomes red.
\citet{cormen2009introduction} did not explicitly mention (e) as a rule, however, it is implicitly mentioned in line 23 of RB-Delete-Fixup(T,x) and caption (b) of Figure 13.7 in \citep[Chapter~13]{cormen2009introduction}.
}
\label{fig:rbt-delete-case-3} 
\end{figure}

The delete operation may happen at the root node, an internal node, or a leaf node. Firstly, if the to-be-deleted node is of degree 2, its value is replaced by the greatest value in the left subtree or the smallest value in the right subtree, transferring the deletion to a degree-1 node or a leaf node. Then, the actual deletion is performed according to the following rules:
\begin{enumerate}
\item \textbf{Deleting a degree-1 node:}
Since degree-1 nodes do not possess a child on one side, the existence of a black node further down their subtree is precluded (since it would violate the 3rd property of RB trees). 
Also, since a node and its child cannot both be red,
it is only possible for a degree-1 node to be a black node with a single red child.
In this case, the value of the red child node is copied to the degree-1 node, and the red child node is deleted.
\item \textbf{Deleting a red leaf node:}
In this case, the node is simply removed and the resulting tree is a legitimate RB tree.
\item \textbf{Deleting a black leaf node:}
After deleting a black leaf node, the number of black nodes from the root node to the leaves of the left and right subtrees of the parent of the deleted node would be different, and the 3rd property of Definition~\ref{def:rbt} would be violated. In this case, until at least one of the rules of Figure~\ref{fig:rbt-delete-case-3} is applicable, the fix-up operations are continued.
\end{enumerate}
The main problem with the rules of Figure~\ref{fig:rbt-delete-case-3} is not their number, but their unclear rationale. For example, the rule of Figure~\ref{fig:rbt-delete-case-3-4} states that if the root of the deficient subtree is black, its sibling is black, and the right child of the sibling is red, then make the right child of the sibling black, and perform a left rotation on the sibling. From an educational point of view, the problem with this rule is that one has no idea what the rationale behind it is.

\section{Left-Leaning Red-Black (LLRB) trees }
\label{sec:llrb}
\begin{figure}[h]%
\centering
\subfloat[]{\includegraphics[scale=\s, valign=c]{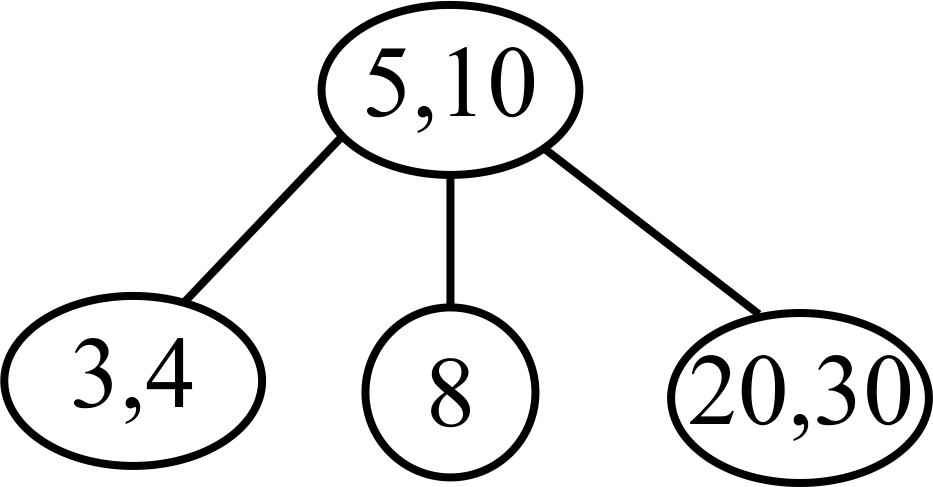}}\qquad\qquad\qquad
\subfloat[]{\includegraphics[scale=\s, valign=c]{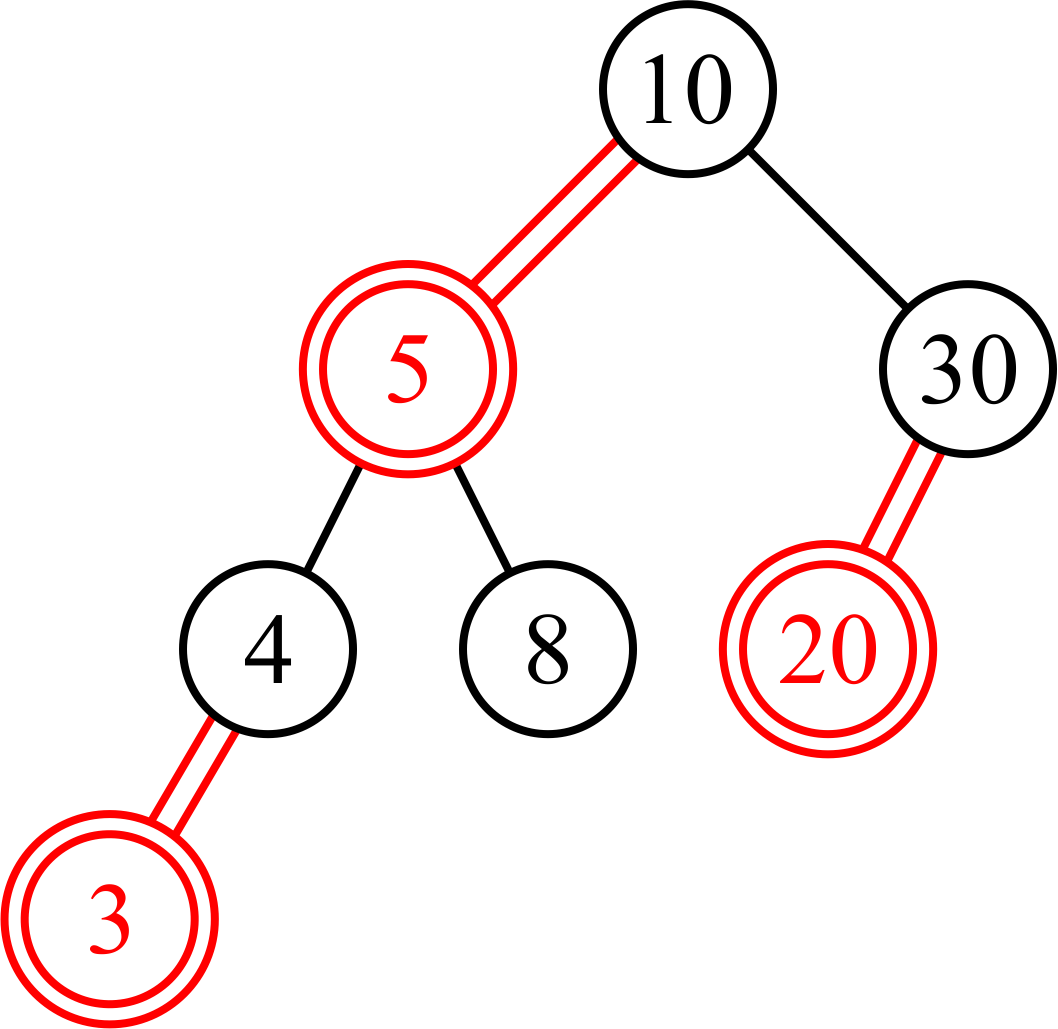}}
\caption{ A 2-3 tree (a) and its equivalent LLRB tree (b).}
\label{fig:23-llrb-equivalence}
\end{figure}

For pedagogical purposes, \citet{sedgewick2008left} proposed 
LLRB trees to lessen the complexity of classical red-black trees. An LLRB tree is a red-black tree in which all red nodes are left children of their parents. LLRB trees have a one-to-one correspondence with 2-3 trees. Figure~\ref{fig:23-llrb-equivalence} shows an example of this one-to-one correspondence. 
\citet{sedgewick2008left} proposed a neat insertion algorithm and taught it in his MOOC algorithms course on Coursera \citep{sedgewickcourse}.  
However, as we will show, the deletion algorithm of LLRB is neither efficient nor suitable for educational purposes. 

\subsection{Insertion algorithm of LLRB trees}
\begin{figure}[h]%
\centering
\subfloat[Left Rotation]{\includegraphics[scale=\s, valign=c]{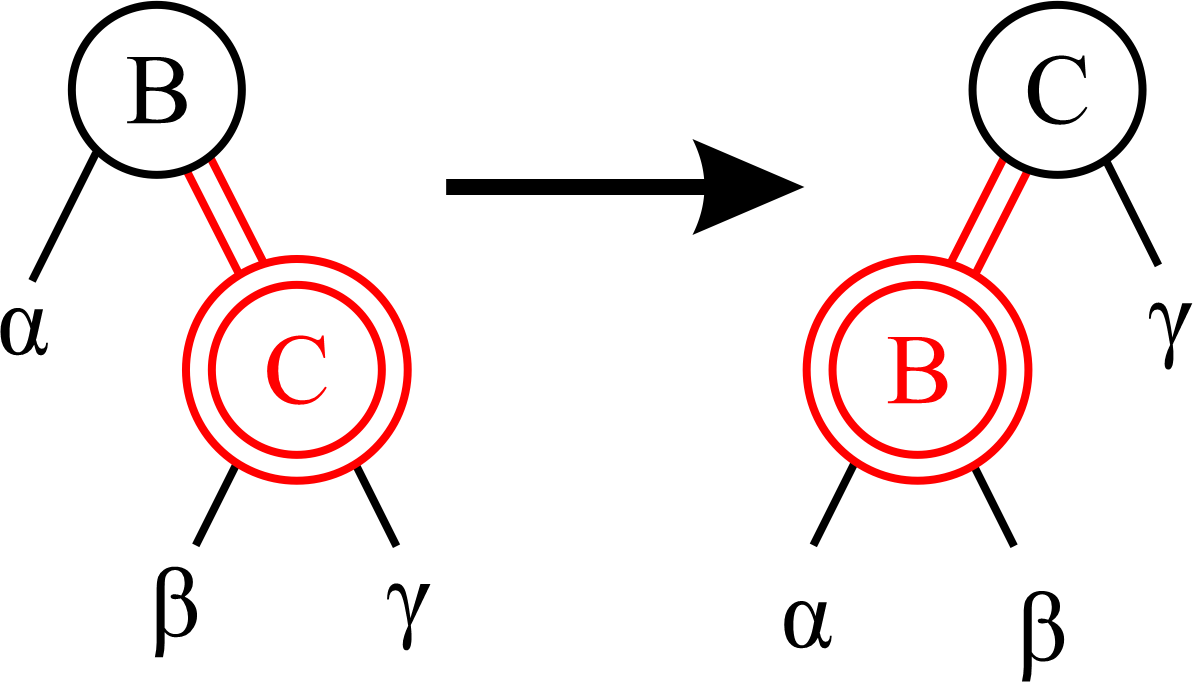}}\qquad
\hspace{1cm}
\subfloat[Right Rotation]{\includegraphics[scale=\s, valign=c]{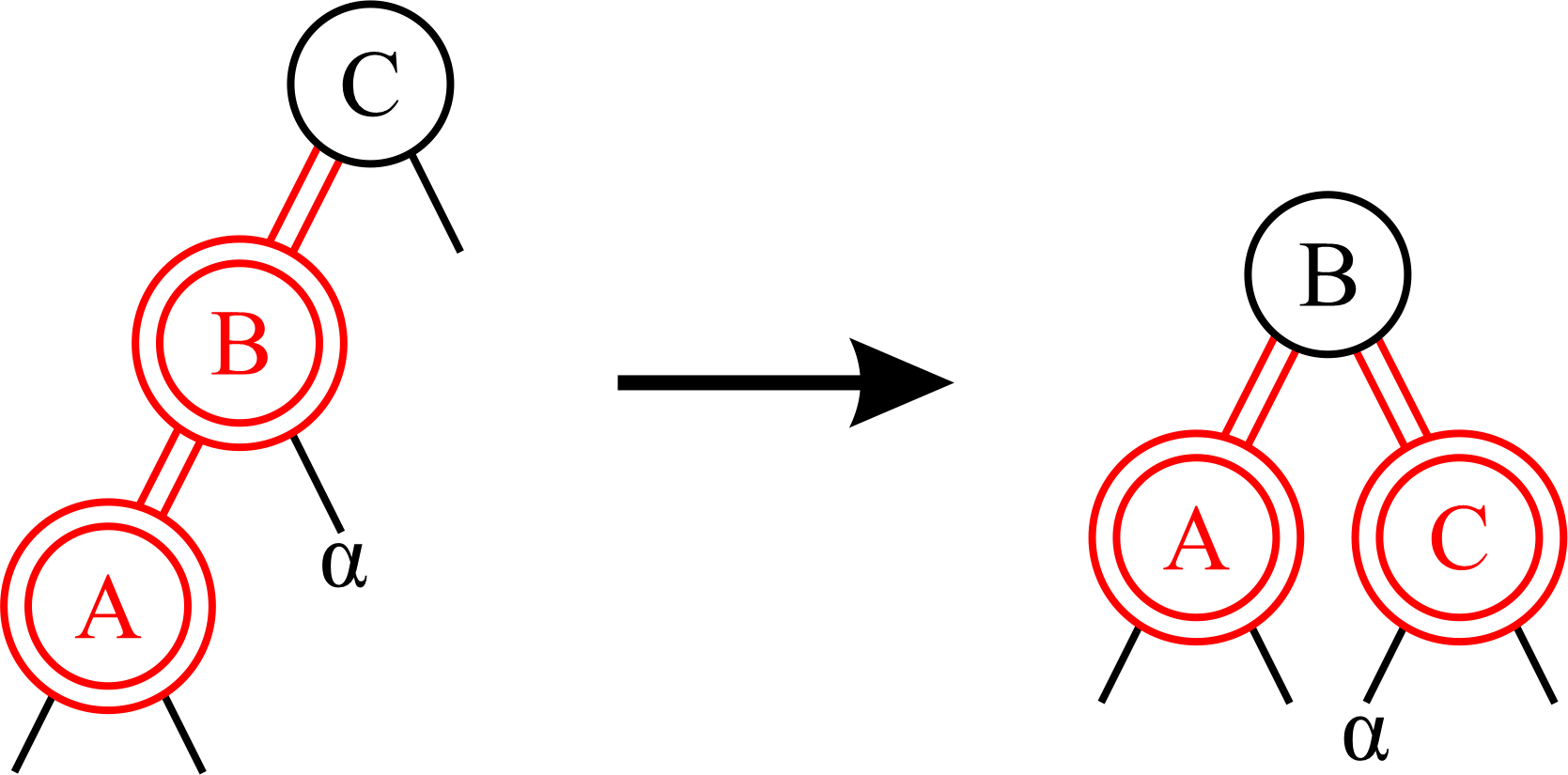}}\qquad
\hspace{1cm}
\subfloat[Color Flip]
{\includegraphics[scale=\s, valign=c]{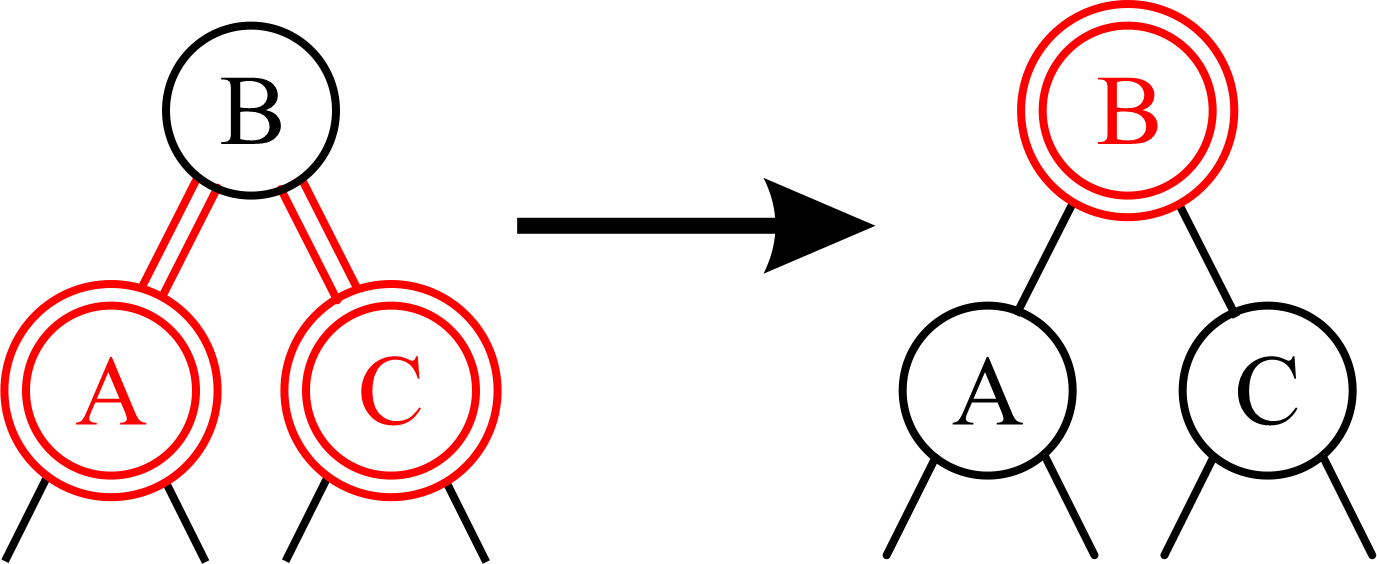}}
\caption{Basic operations of LLRB trees to fix up a tree after an insertion. Note that the rules for applying the left and right rotations are not completely symmetric. The right rotation is applied when two consecutive left children are red, while the left rotation is applied on a right red child.}
\label{fig:llrb-insert}
\end{figure}

As in classical RB trees, the \algname{insert} algorithm of LLRB trees starts by inserting a new leaf node into a binary search tree with the color red.
In addition to the possibility of the violation of the 2nd property of RB trees in Definition~\ref{def:rbt}, the inserted node could be a right child, violating the sole new constraint of LLRB trees.
\citet{sedgewick2008left} proposed the three operations of left rotation, right rotation, and color flip to transform the resulting tree into a correct LLRB tree (Figure~\ref{fig:llrb-insert}).
Note that, in contrast to classical RB trees, where there were three other symmetric cases, since LLRB trees do not permit red right children, here all the cases are the three ones shown in Figure~\ref{fig:llrb-insert}.
One of the important weaknesses of the \algname{insert} algorithm of LLRB is that these rules should be applied until the root node is reached, even though it is possible to infer that the tree has been fixed up long before reaching the root. The reason for this inefficiency is that the \algname{insert} algorithm is implemented recursively and there is no way to empty the call stack except throwing an exception. In fact, our attempt to modify the code of LLRB to terminate the fix-up operation by throwing an exception led to a severe slowdown of the algorithm.

\subsection{Deletion algorithm of LLRB tree}
\begin{figure}[h]%
\centering
\includegraphics[scale=\s, valign=c]{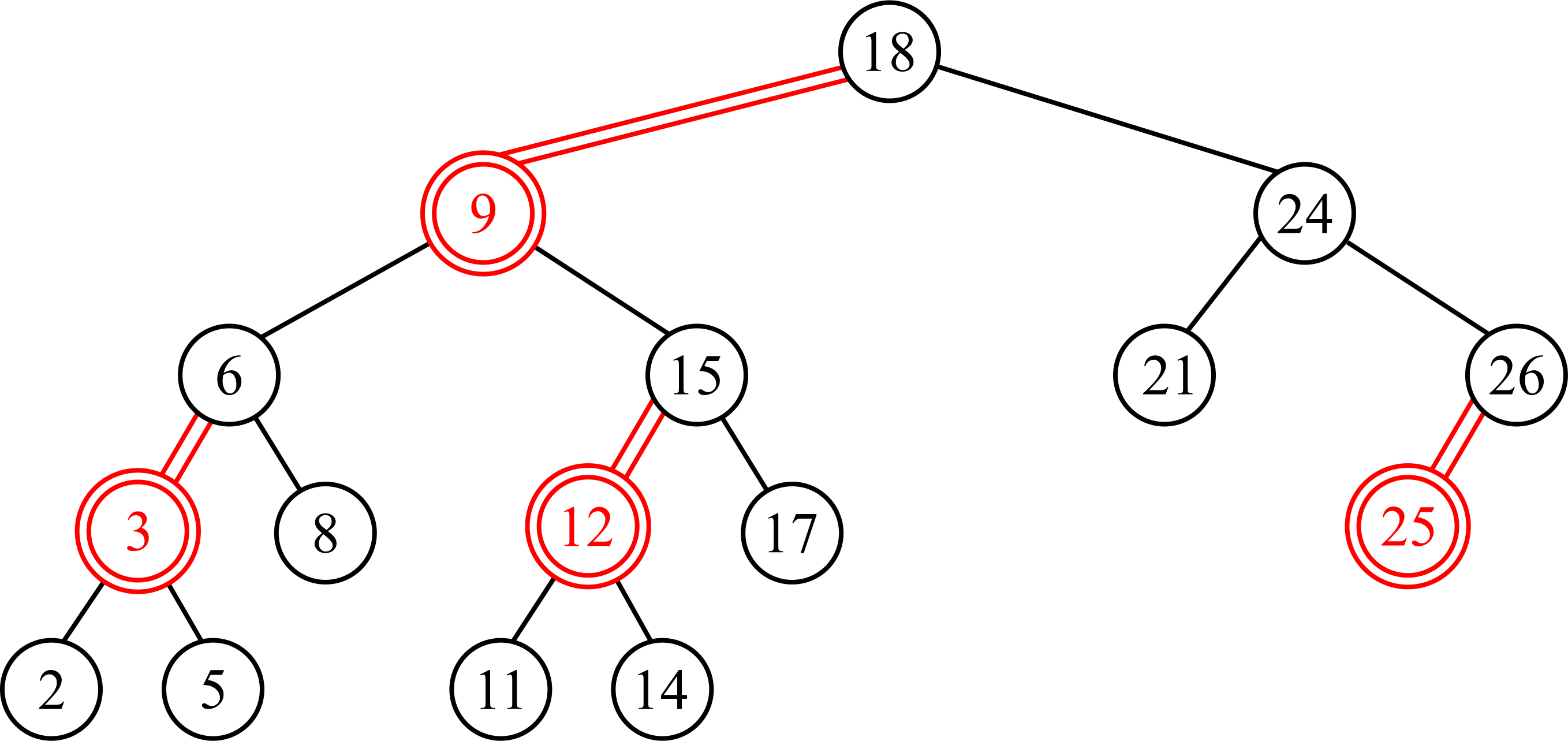}
\caption{An example showing the inefficiency and complexity of the delete operation in LLRB. Deletion of 25, in the top-down pass, leads to the sequence of operation: Right-Rotate(18), Color-Flip(18), Right-Rotate(18), Color-Flip(15). Then the algorithm deletes 25 and continues with Left-Rotate(18) and Left-Rotate(9) when returning from recursive calls. This is while, in this example, 25 is a red node that could be simply deleted without any rotations or color flips.
}
\label{fig:delete-llrb}
\end{figure}

\citet{sedgewick2008left} proposed a recursive top-down algorithm for deletion in LLRB trees. 
To delete a node, the algorithm starts from the root node and moves left/right towards the to-be-deleted node. 
The algorithm prepares the scene to apply the actual deletion to a red node and, therefore, as it descends the tree it ensures that either the current node or one of its children is red. If it is not the case, the algorithm enforces this property by two methods named "moveRedLeft" and "moveRedRight".
As the deletion algorithm descends the tree, it modifies the tree extensively and causes immense changes.
This is very inefficient since it is possible that the query node does not exist, or it is already red and, therefore, can be simply deleted.  Figure~\ref{fig:delete-llrb} shows an example of a tree in which the deletion operation is as simple as solely deleting the node with the given key, while the \algname{delete} algorithm of LLRB engages in immense modifications to the tree. For a complete example showing the functionality of the \algname{delete} algorithm of LLRB trees see \url{https://profsite.um.ac.ir/~k.ghiasi/publications/PS-RBT/llrb-delete-images.pdf}.

\section{The considered framework: 2-3 RB trees}
\label{sec:2-3 RB}
\begin{figure}[h]%
\centering
\subfloat[]{\includegraphics[scale=\s, valign=c]{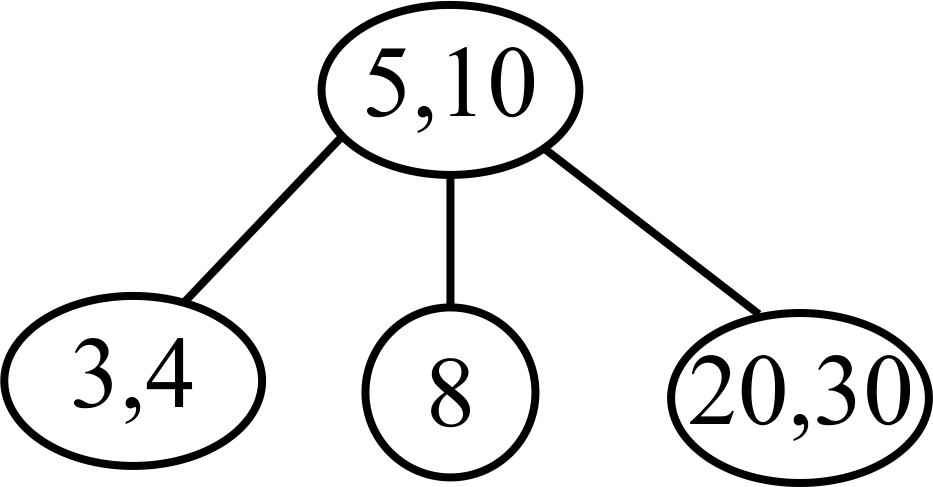}}\qquad
\subfloat[]{\includegraphics[scale=\s, valign=c]{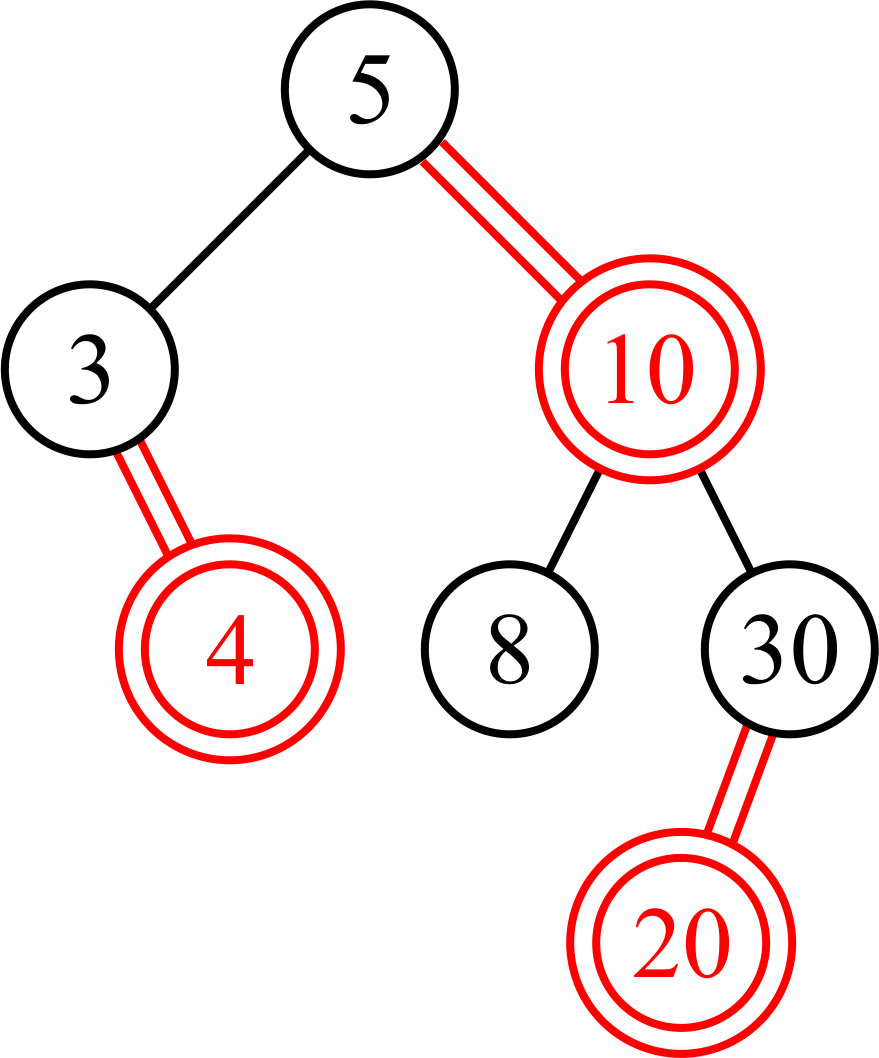}}\qquad
\subfloat[]
{\includegraphics[scale=\s, valign=c]{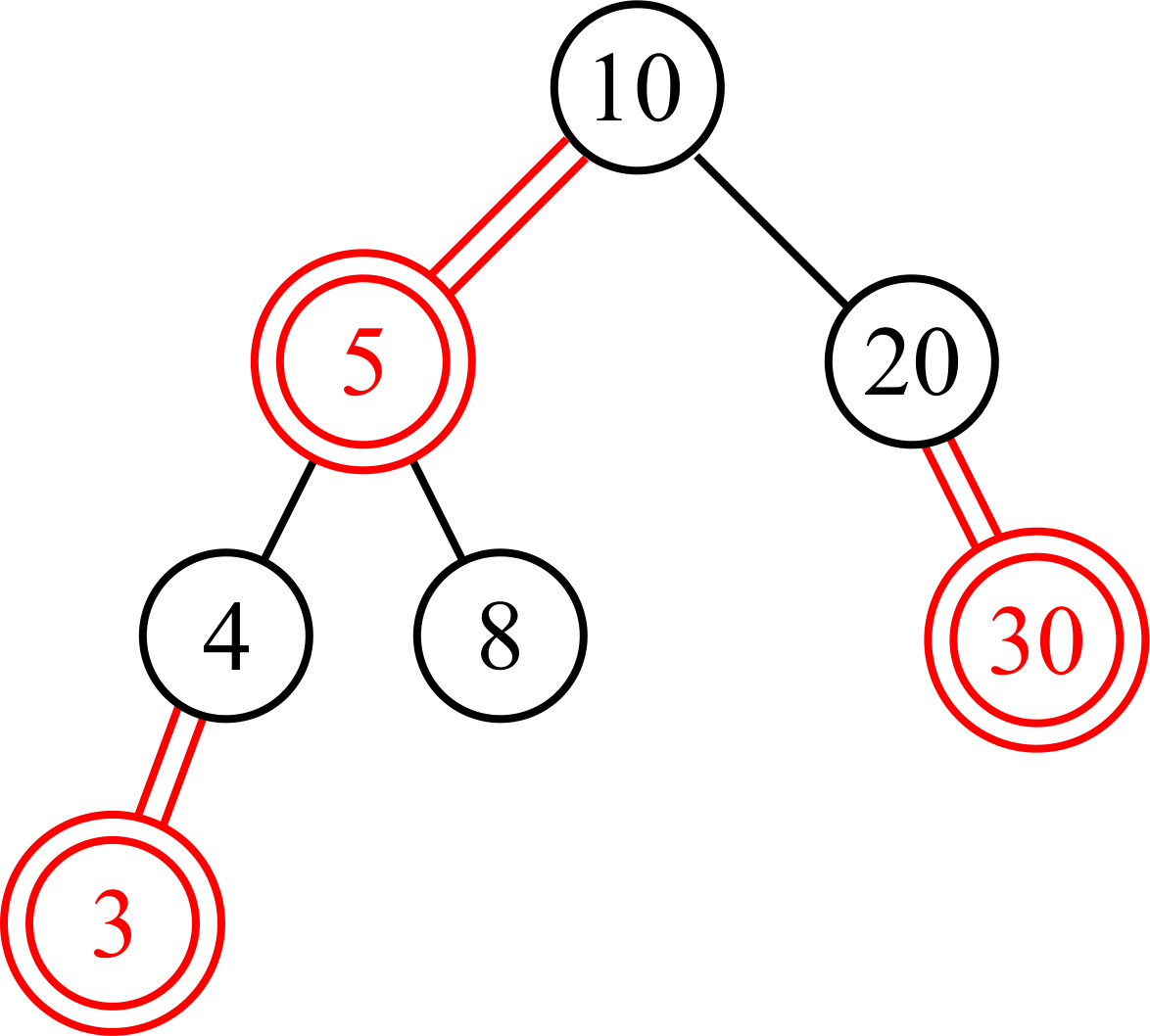}}
\caption{(a) A 2-3 tree and (b,c) two of its equivalent 2-3 RB trees.}
\label{fig:2-3 RB-23-equiv}
\end{figure}

We define a 2-3 RB tree as a red-black tree in which both children of a node cannot be red. Note that, like \citep{bayer1972symmetric} and in contrast to \citep{bayer1971binary, andersson1993balanced, sedgewick2008left}, 2-3 RB trees treat the left and right children symmetrically. 
While LLRB trees are in one-to-one correspondence with 2-3 trees, there might be multiple equivalent 2-3 RB trees for a given 2-3 tree. 
Figure~\ref{fig:2-3 RB-23-equiv} illustrates a 2-3 tree and two of its equivalent 2-3 RB trees.

\subsection{Insertion algorithm for 2-3 RB trees}
\begin{figure}[h]%
\centering
\subfloat[Ensure having aligned red links.]{
\includegraphics[scale=\s, valign=c]{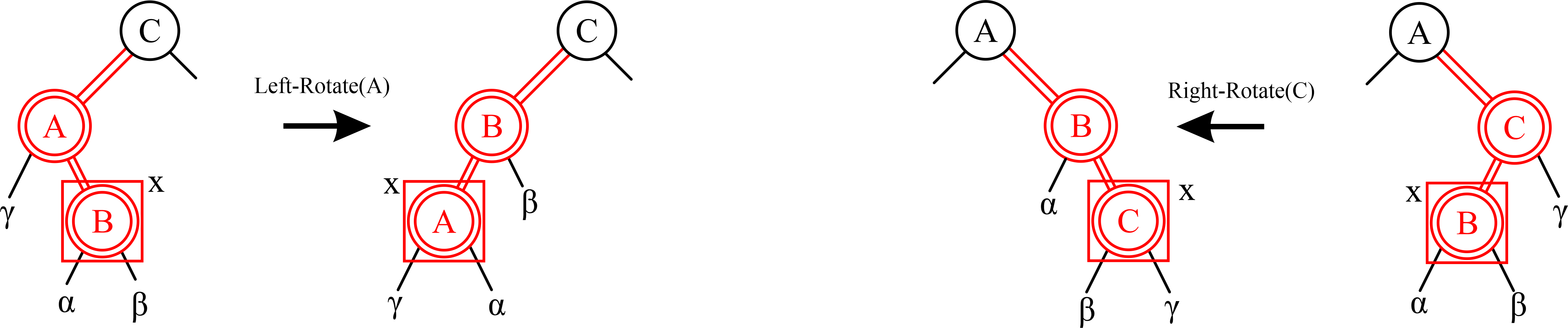}
\label{fig:2-3 RB-insert-case-I}
}
\\
\vspace{0.4cm}
\subfloat[Convert red parent-child pairs to red siblings.]{
\includegraphics[scale=\s, valign=c]{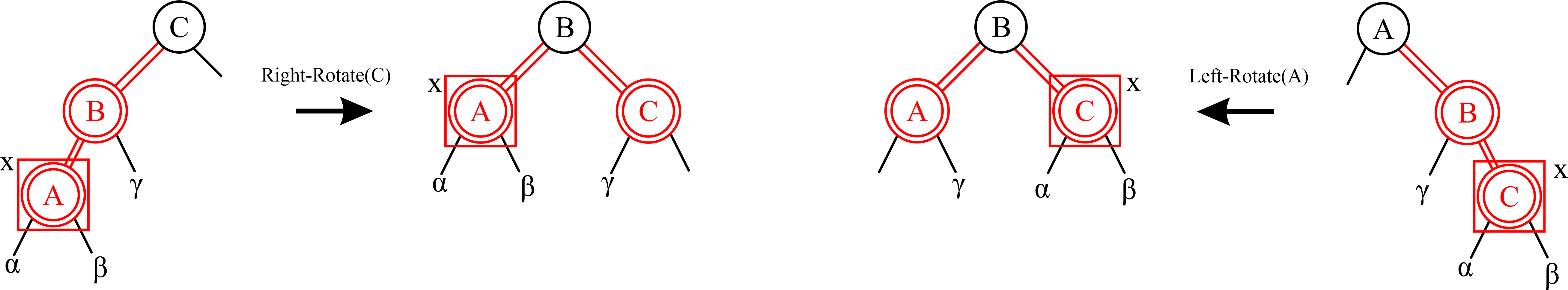}
\label{fig:2-3 RB-insert-case-II}
}
\\
\vspace{0.4cm}
\subfloat[Flip color on red siblings.]{
\includegraphics[scale=\s, valign=c]{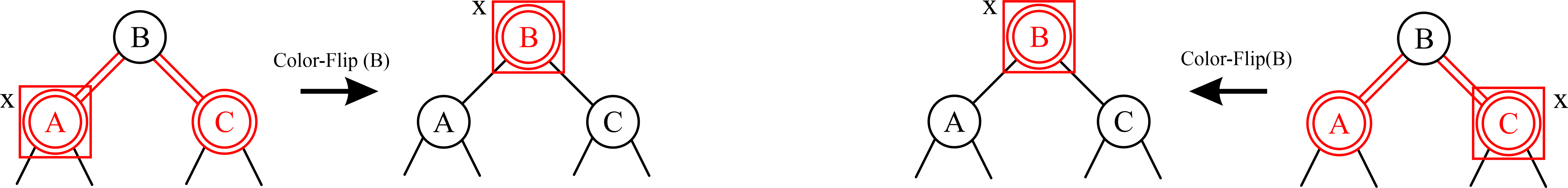}
\label{fig:2-3 RB-insert-case-III}
}
\caption{Rules for fixing 2-3 RB trees after inserting a new node. The rules are applied recursively. The node designated with letter $x$ and a square is the node for which the rules are matched. Initially, $x$ is the inserted node. (a) and (b) show the rules when a node and its parent are both red. (c) shows the rules when a node and its sibling are both red. 
Note that, although the number of rules are twice that of LLRB, the simplicity and intuitiveness of the rules are the same.
}
\label{fig:2-3 RB-insert} 
\end{figure}

\begin{figure}[H]
\centering
\includegraphics[scale=\s, valign=c]{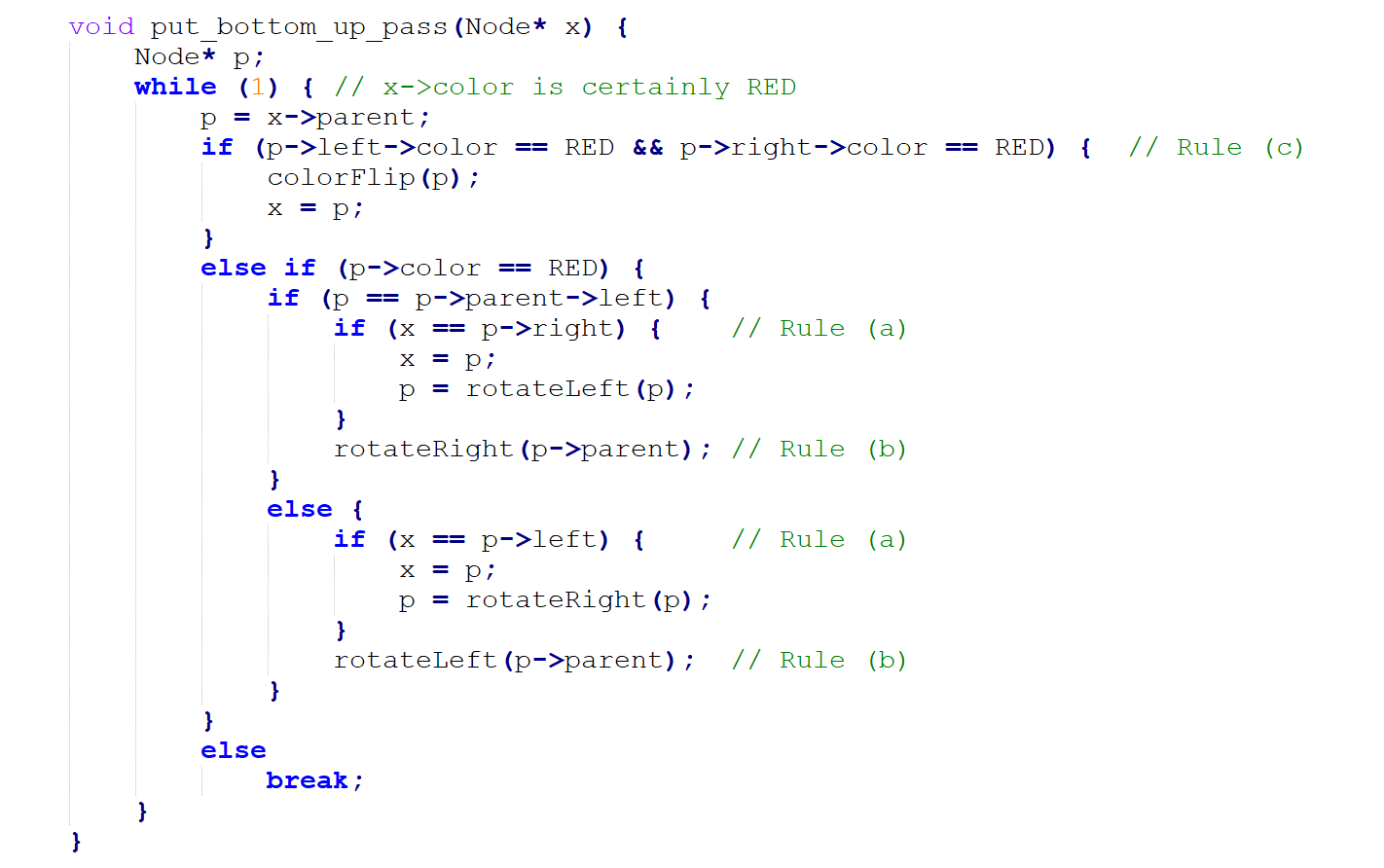}
\caption{The code fragment implementing the rules of Figure~\ref{fig:2-3 RB-insert} for fixing the 2-3 RB tree after insertion of a new node. The method receives a pointer to the newly inserted node.}
\label{fig:2-3-insert-code}
\end{figure}

To insert a value in a 2-3 RB tree, we initially insert it with the color red in the position determined by the rules of binary search trees.
We denote the newly inserted node by $x$. 
Inductively, we denote the sole red node that might have a red parent or a red sibling by $x$.
We preserve this property that, if there is a violation of some property of 2-3 RB trees, it should either be that $x$ and its parent are red (case I) or $x$ and its sibling are red (case II).
We perform fix-up operations until we obtain a legitimate 2-3 RB tree. 
Our proposed rules for case I, in which the node $x$ and its parent are both red, are shown in Figure~\ref{fig:2-3 RB-insert-case-I} and Figure~\ref{fig:2-3 RB-insert-case-II}.
In case II, in which $x$ and its sibling are red, we propose a color-flip operation as shown in Figure~\ref{fig:2-3 RB-insert-case-III}. 
We terminate the fix-up operations as soon as none of the rules of Figure~\ref{fig:2-3 RB-insert} applies. 
Finally, we reset the color of the root to black, to ensure that this fix-up process has not changed its color to red. 
Figure~\ref{fig:2-3-insert-code} shows the implementation of these rules in C++.
\begin{proposition}
The fix-up operations of the \algname{insert} algorithm of 2-3 RB trees terminate.
\end{proposition}
\proof As is clear from Figure~\ref{fig:2-3 RB-insert}, at each step, the node marked with $x$ becomes one level closer to the root node. Therefore, the maximum possible number of fix-up operations is the height of the tree.


\subsection{The proposed parity-seeking \algname{delete} algorithm for 2-3 RB trees}
\begin{figure}[H]%
\centering
\subfloat[]{
\includegraphics[scale=\s, valign=c]{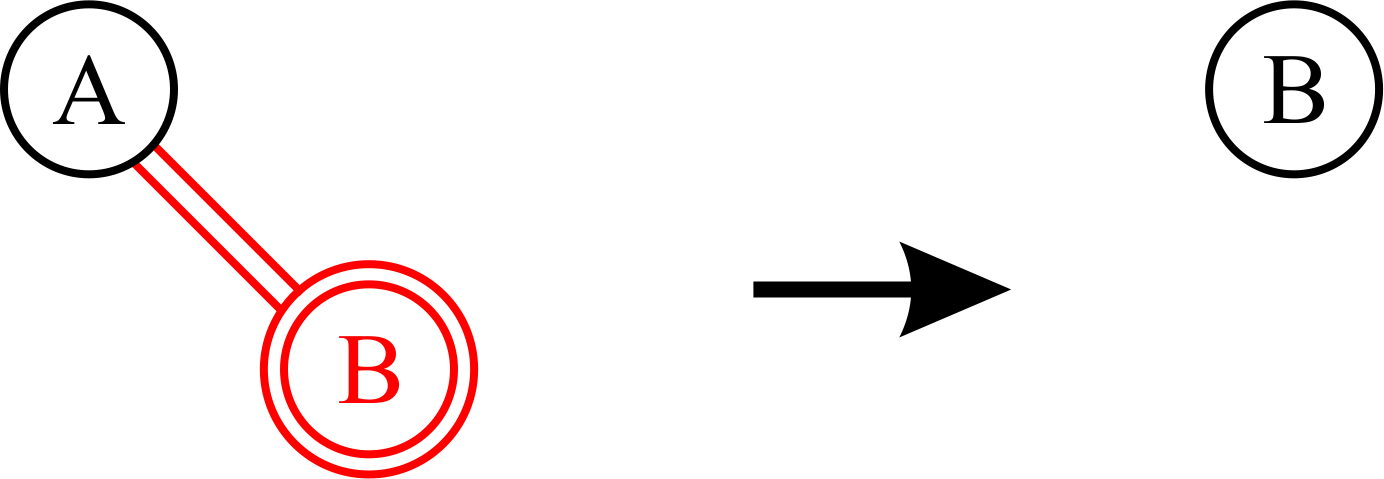}
\label{fig:2-3 RB-delete-degree-one-1}
}
\qquad\qquad\qquad\qquad
\subfloat[]{
\includegraphics[scale=\s, valign=c]{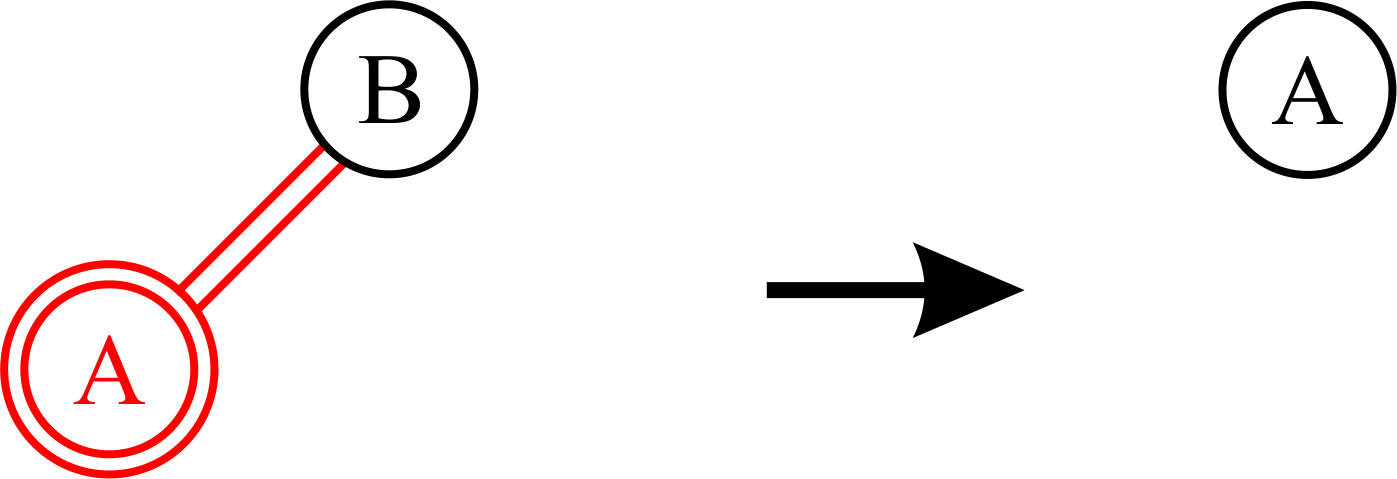}
\label{fig:2-3 RB-delete-degree-one-2}
}
\caption{If the target of deletion is a degree-1 one, its content is substituted by the content of its child, and the child is deleted.}
\label{fig:2-3 RB-delete-degree-one}
\vspace{0.5cm}
\subfloat[]{
\includegraphics[scale=\s, valign=c]{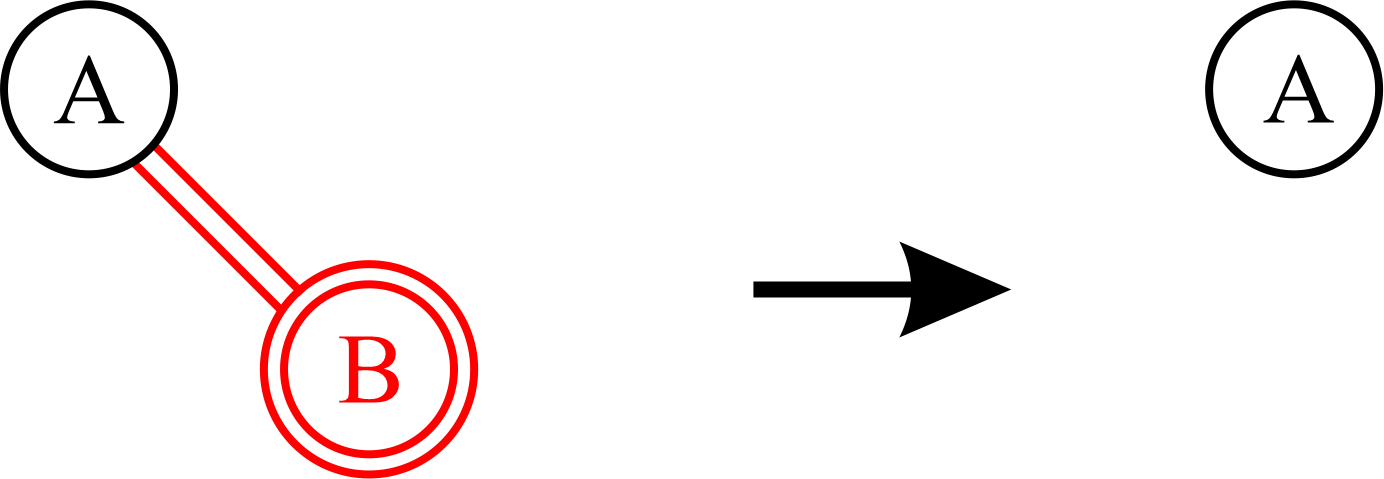}
\label{fig:2-3 RB-delete-red-leaf-1}
}
\qquad\qquad\qquad\qquad
\subfloat[]{
\includegraphics[scale=\s, valign=c]{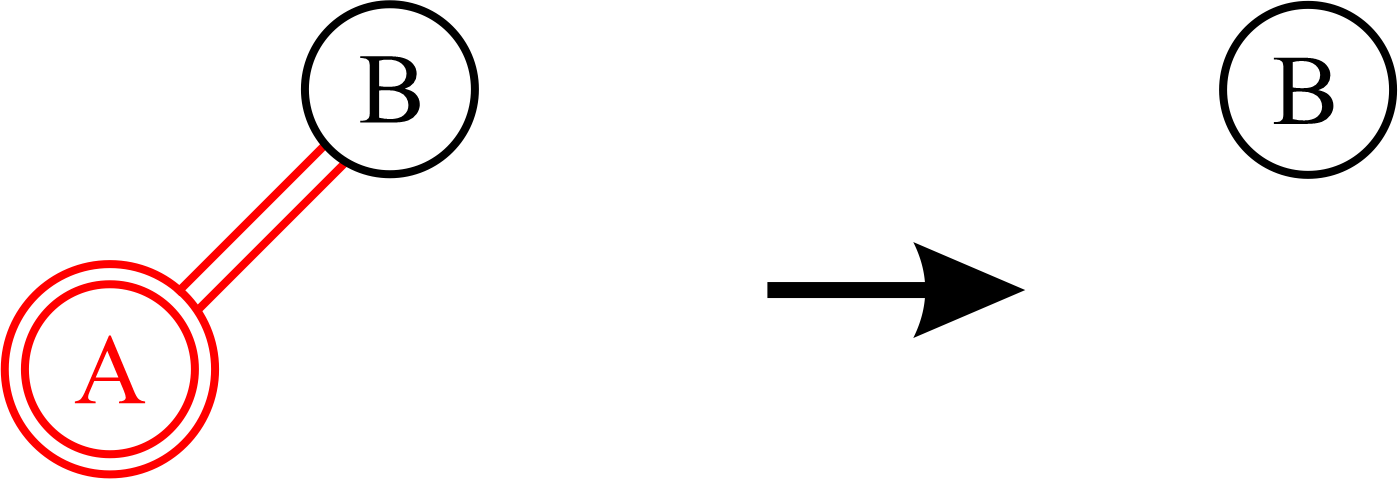}
\label{fig:2-3 RB-delete-red-leaf-2}
}
\caption{If the target of deletion is red, it is simply removed, and a valid 2-3 RB tree is obtained.}
\label{fig:2-3 RB-delete-red-leaf}
\end{figure}

In this section, we describe our proposed parity-seeking \algname{delete} algorithm in the context of 2-3 RB trees.
Firstly, according to the deletion rules of binary search trees, the initial delete operation is transferred to a leaf or a degree-1 node. 
Now, if the degree of the to-be-deleted node is one, then, from property 3 of Definition~\ref{def:rbt}, it follows that its whole subtree is a single red child. Therefore, to delete a degree-1 node, it suffices to delete its red child and put its value in its parent (Figure~\ref{fig:2-3 RB-delete-degree-one}).
Now, consider the case of deleting a leaf node. 
If the leaf node is red, then it can be simply deleted and the resulting tree is a valid 2-3 RB tree (Figure~\ref{fig:2-3 RB-delete-red-leaf}).

The hard case is deleting a black leaf node. 
First, let us define deficient subtrees.
\begin{definition}[Deficient subtree]
A subtree rooted at a node $x$ is deficient if (1) assuming $x$ is black, it is a 2-3 RB tree, and (2) the number of visited black nodes from $x$ to the leaves is one less than that of $x$’s sibling.
\end{definition}

Assume that we want to delete a black leaf node named $z$. 
After deleting $z$, we replace it with \textit{nilSentinel} and set the parent of \textit{nilSentinel} to the parent of $z$. Therefore, initially, the root of the deficient subtree is the \textit{nilSentinel} node.
Inductively, assume that $x$ is the root of the deficient subtree, and $y$ is its sibling. 
Our parity-seeking \algname{delete} algorithm works as follows: it either fixes the deficiency of the node $x$ or also makes its sibling $y$ deficient, elevating the deficiency to the parent node.
There are three possibilities: 
\begin{enumerate}[label=\Roman*.]
\item $x$ is red.
\item $x$ and $y$ are both black.
\item $x$ is black and $y$ is red.
\end{enumerate}
Case I is simply handled by changing the color of $x$ to black, which resolves the deficiency of $x$. The rule corresponding to this case is shown in Figure~\ref{fig:2-3 RB-delete-main-1}. 

Now consider case II where both the root of the deficient subtree, viz. $x$, and its sibling, viz. $y$, are black.
We move the deficiency one level higher by turning $y$ red. The rule corresponding to this case is shown in Figure~\ref{fig:2-3 RB-delete-main-2}. Please note that there is no special handling for the case that the whole tree becomes deficient, as it is automatically handled by rules (a) and (b) of Figure~\ref{fig:2-3 RB-delete-main}.
If one of y’s children is red, turning $y$ red according to this rule would violate the 2nd property in Definition~\ref{def:rbt}.
We have devised separate fixing rules for this situation that would be explained in subsection~\ref{sec:ps-delete-fixing-rules}. 
These fixing rules not only reinstate the 2nd property of Definition~\ref{def:rbt} but also resolve the deficiency problem altogether.

\begin{figure}[H]
\centering
\includegraphics[scale=\s, valign=c]{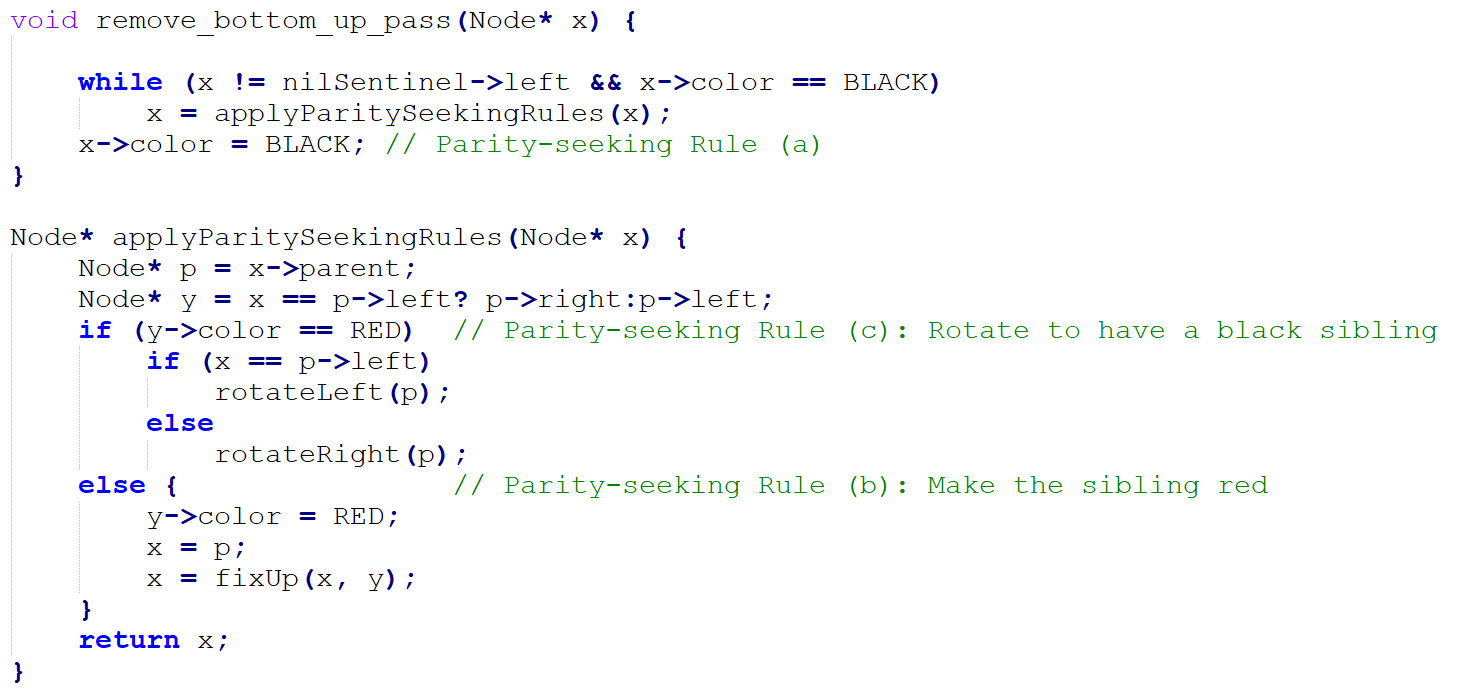}
\caption{Implementation of the main parity-seeking rules}
\label{fig:2-3 RB-delete-main-code}
\end{figure}

Finally, consider case III where the root of the deficient subtree, viz. $x$ is black, and its sibling, viz. $y$, is red.
Since $y$ is red, children of $y$ are black. 
We can neither fix the deficiency of $x$ as $x$ is black nor make the sibling deficient since $y$ is red.
We perform a rotation on the common parent of $x$ and $y$ so that the new sibling of $x$ becomes one of the children of $y$. Since the new sibling of x is black, the algorithm returns to case II. Figure~\ref{fig:2-3 RB-delete-main-3} illustrates this situation.
Figure~\ref{fig:2-3 RB-delete-main-code} shows the implementation of the main rules of the parity-seeking algorithm in C++.

\subsubsection{Fixing rules}
\label{sec:ps-delete-fixing-rules}
\begin{figure}[H]
\centering
\includegraphics[scale=\s, valign=c]{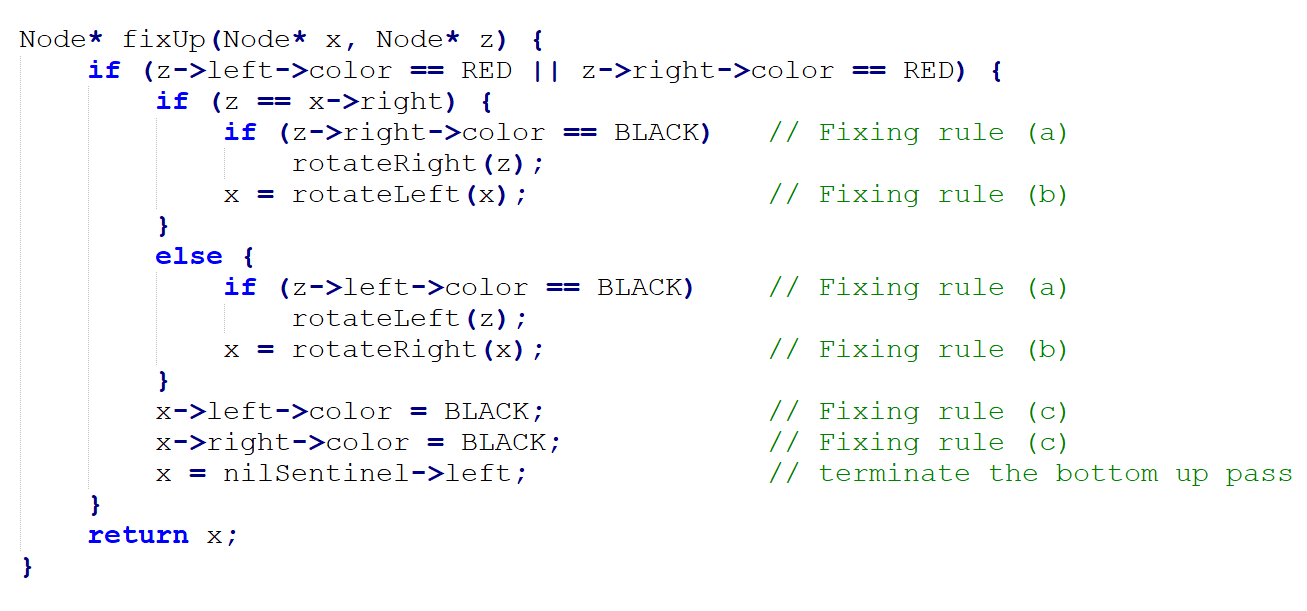}
\caption{Implementation of the fixing rules of the parity-seeking algorithm.}
\label{fig:2-3 RB-delete-fixing-code}
\end{figure}

After applying rule (b) of the parity-seeking deletion algorithm, as shown in Figure~\ref{fig:2-3 RB-delete-main-2}, a node, which we denote by $z$, becomes red. 
Consequently, if one of the children of $z$ has already been red, then the second property of RB trees is violated. This situation is resolved by the 'fixing' rules shown in Figure~\ref{fig:2-3 RB-delete-black-leaf-fix-double-red}. 
Figure~\ref{fig:2-3 RB-delete-fixing-code} shows the implementation of these fixing rules in C++.

\subsubsection{Proving termination and correctness}
In contrast to the \algname{insert} algorithm in which the considered node was steadily moving up the tree, in the \algname{delete} algorithm the deficient subtree can both move up or down the tree. In the following proposition, we prove that, despite this, the \algname{delete} algorithm of 2-3 RB trees terminates.

\begin{figure}[H]%
\centering
\subfloat[If you can fix the deficiency of the current subtree, then do it.]{
\includegraphics[scale=\s, valign=c]{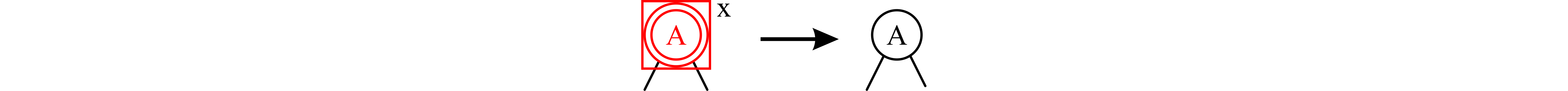}
\label{fig:2-3 RB-delete-main-1}
}\\
\vspace{0.5cm}
\subfloat[If you cannot fix the deficiency of the current subtree, but you have a black sibling, then change the color of the sibling to red to become on par. $z$ denotes the sole red child of $x$ which is used in the subsequent fixing rules of Figure~\ref{fig:2-3 RB-delete-black-leaf-fix-double-red}.]{
\includegraphics[scale=\s, valign=c]{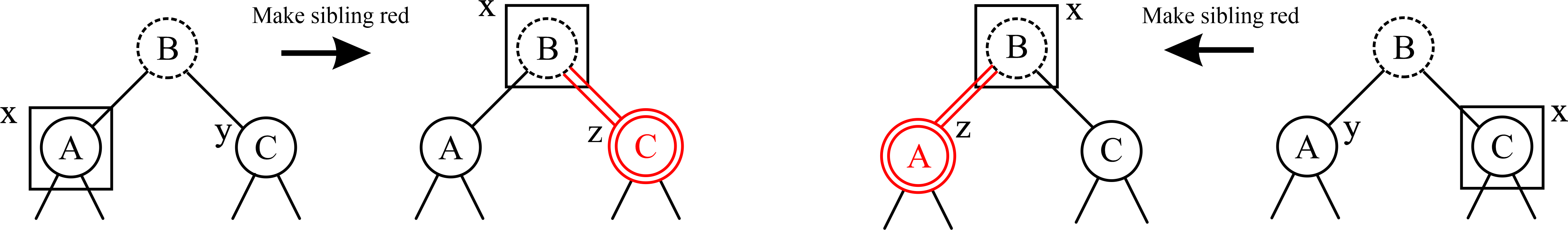}
\label{fig:2-3 RB-delete-main-2}
}\\
\vspace{0.5cm}
\subfloat[If none of the previous rules can be applied, provide a black sibling by an appropriate rotation on the parent.]{
\includegraphics[scale=\s, valign=c]{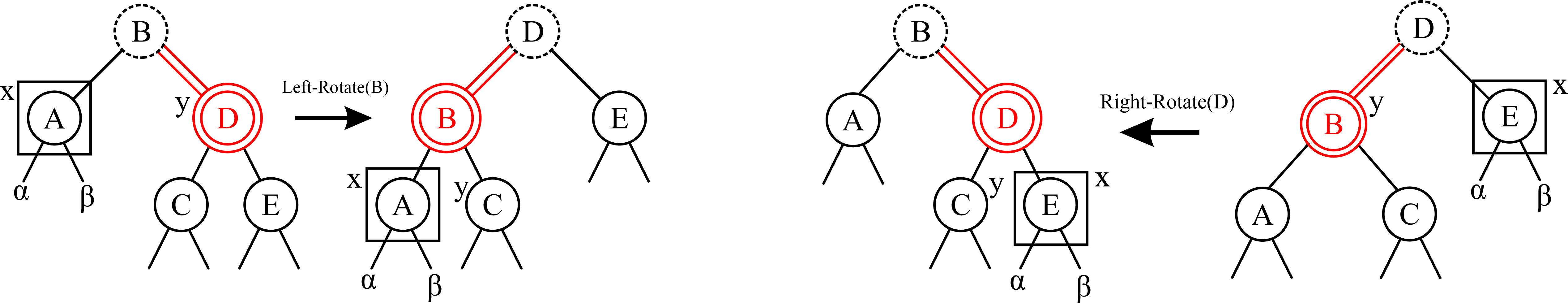}
\label{fig:2-3 RB-delete-main-3}
}
\caption{Main rules for the case of deleting a black leaf node from 2-3 RB trees. After deleting a black leaf node, its subtree becomes deficient. We have denoted the root of the deficient subtree by $x$ and its sibling by $y$. The rationale behind these rules is very simple: either fix the deficiency, or make the sibling deficient as well, lifting up the deficiency to the parent node.
(a) Case I in which $x$ is red. The deficiency is simply solved by making $x$ black.
(b) Case II in which both $x$ and $y$ are black. 
The solution is to make $y$ red and lift up the deficiency to the parent of $x$. (c) Case III in which $x$ is black and $y$ is red. The solution is to perform a rotation so that the new sibling of $x$ becomes black. Since $y$ is red, $y$'s children are certainly black and, therefore, the new sibling of $x$ would be black.
This returns to case II in which the sibling of $x$ is black.}
\label{fig:2-3 RB-delete-main}
\end{figure}

\begin{figure}[H]%
\centering
\subfloat[If red links are not aligned, rotate to have aligned red links.]{
\includegraphics[scale=\s, valign=c]{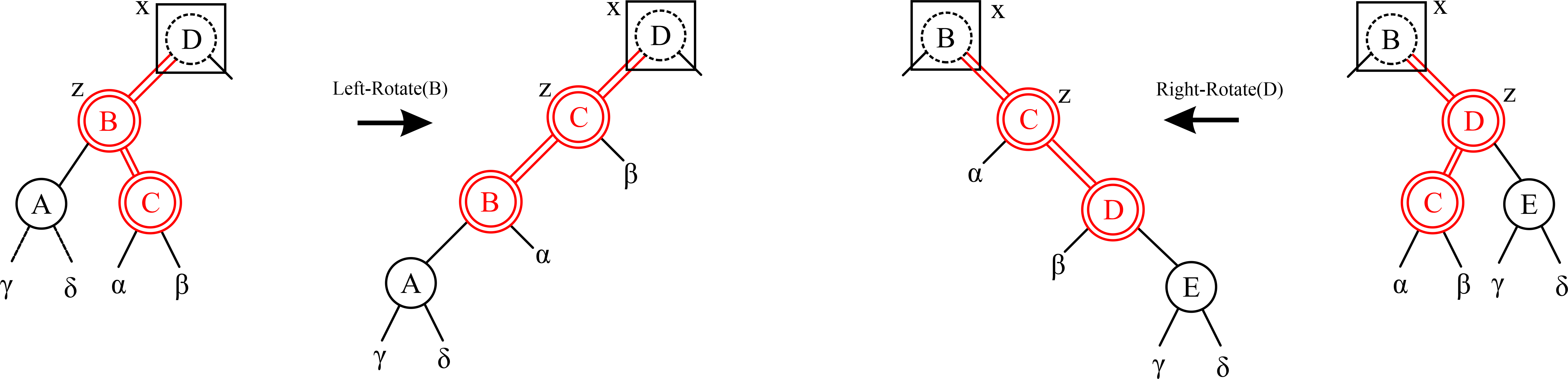}
}
\\
\vspace{0.5cm}
\subfloat[Rotate aligned red links to have two red children.]{
\includegraphics[scale=\s, valign=c]{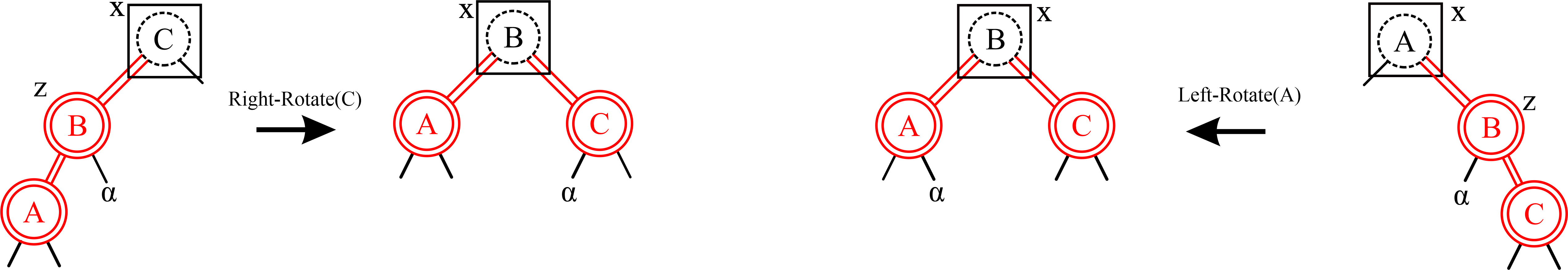}
}
\\
\vspace{0.5cm}
\subfloat[Flip color of children.]{
\includegraphics[scale=\s, valign=c]{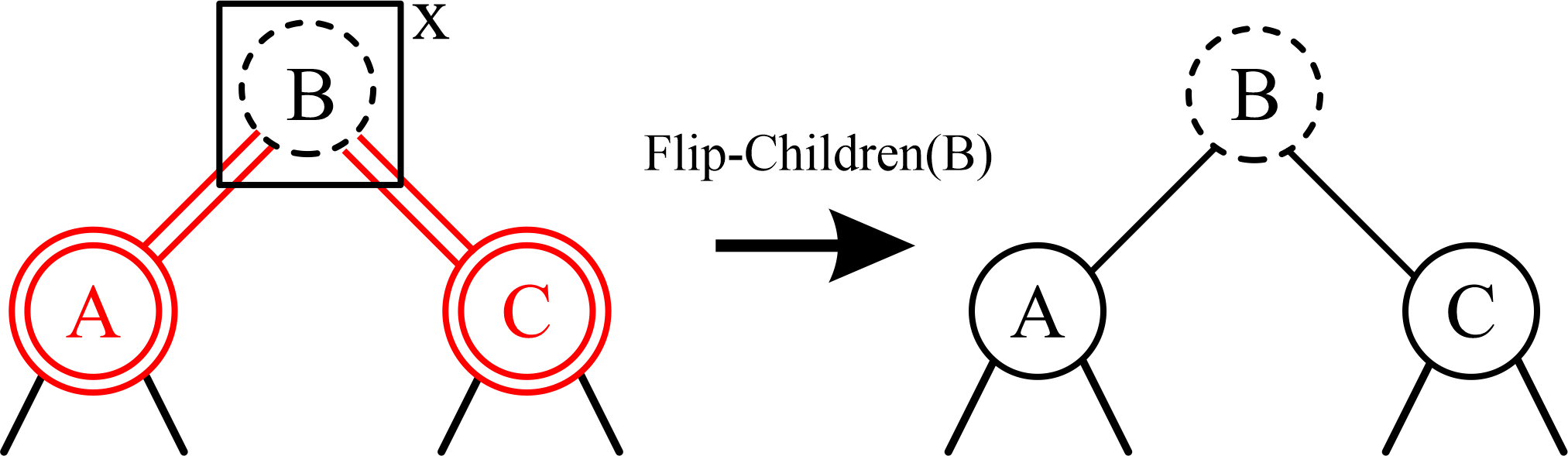}
}
\caption{The fixing rules for the case of deleting a black leaf node from 2-3 trees. When applying rule (b) of Figure~\ref{fig:2-3 RB-delete-main}, if one of $y$'s children is red, the 2nd proprty of Definition~\ref{def:rbt} is violated. (a) Parent-child red links are aligned. (b) Re-establishing the 2nd property of Definition~\ref{def:rbt} by performing a rotation on aligned parent-child red links. (c) A deficient node with red children appears which helps to solve the deficiency altogether: the children are color-flipped and the deficiency is resolved. 
}
\label{fig:2-3 RB-delete-black-leaf-fix-double-red}
\end{figure}

\begin{proposition}
The proposed parity-seeking algorithm for deletion in 2-3 RB trees terminates and generates a legitimate 2-3 RB tree.
\end{proposition}
\proof We need to prove that, in all the three cases of the \algname{delete} algorithm, the problem of deficiency is resolved. We have:
\begin{itemize}
\item In case I, where $x$ is red, the deficiency problem is completely resolved by making $x$ black (Figure~\ref{fig:2-3 RB-delete-main-1}). In this case the algorithm clearly terminates.
\item In case II, where both $x$ and $y$ are black, the deficiency moves one level closer to the root node, as shown in Figure~\ref{fig:2-3 RB-delete-main-2}.
Besides, if $y$ has a red child, then the deficiency problem would be completely resolved, as shown in Figure~\ref{fig:2-3 RB-delete-black-leaf-fix-double-red}.
\item In case III, where $x$ is black and $y$ is red, the algorithm eventually moves to case II.
Considering both cases of Figure~\ref{fig:2-3 RB-delete-main-3}, 
if at least one of C’s children were red, the deficiency problem was resolved immediately as was shown in Figures~\ref{fig:2-3 RB-delete-black-leaf-fix-double-red}. On the other hand, if both children of $C$ were black, then, after applying rule (b) of Figure~\ref{fig:2-3 RB-delete-main}, $C$ becomes red and the deficiency problem lifts up to the red node $B/D$.
The deficiency of the red node $B/D$ is then immediately resolved by changing its color to black by rule (a) of Figure~\ref{fig:2-3 RB-delete-main}.
\end{itemize}


\section{A Parity-Seeking  \algname{delete} algorithm for classical RB trees}
\label{sec:clarification}
\begin{figure}[H]%
\centering
\subfloat[Rule (a) of the classical delete algorithm and main rule (c) of the parity-seeking algorithm are the same.]{
\includegraphics[scale=\sw, valign=c]{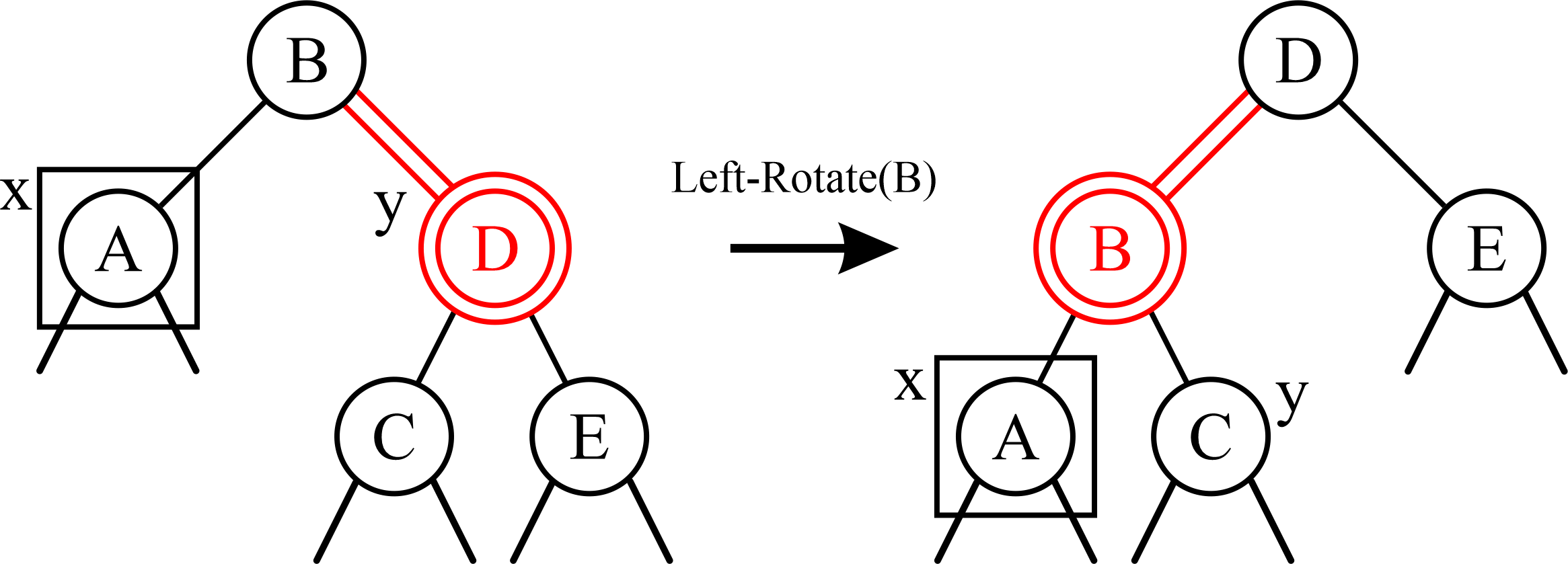}
\label{fig:2-3 RB-delete-case-3}
}
\qquad\qquad
\subfloat[Rule (b) of the classical delete algorithm is handled by main rule (b) of the parity-seeking algorithm.]{
\includegraphics[scale=\sw, valign=c]{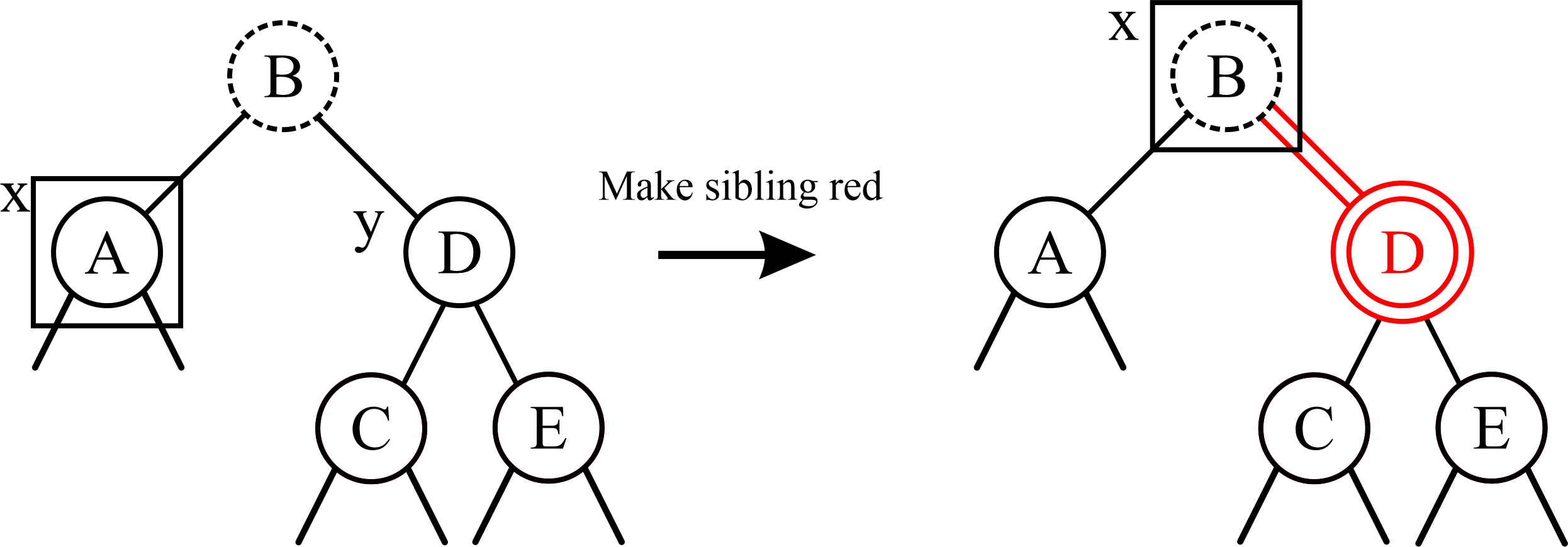}
\label{fig:2-3 RB-delete-case-2-1}
}
\\
\vspace{0.5cm}
\subfloat[This sub-figure shows how the left situation is transformed to the right one using the rules of the parity-seeking delete algorithm. The same result is obtained by successive applications of rules (c) and (d) of the classical delete algorithm.]{
\includegraphics[scale=\sw, valign=c]{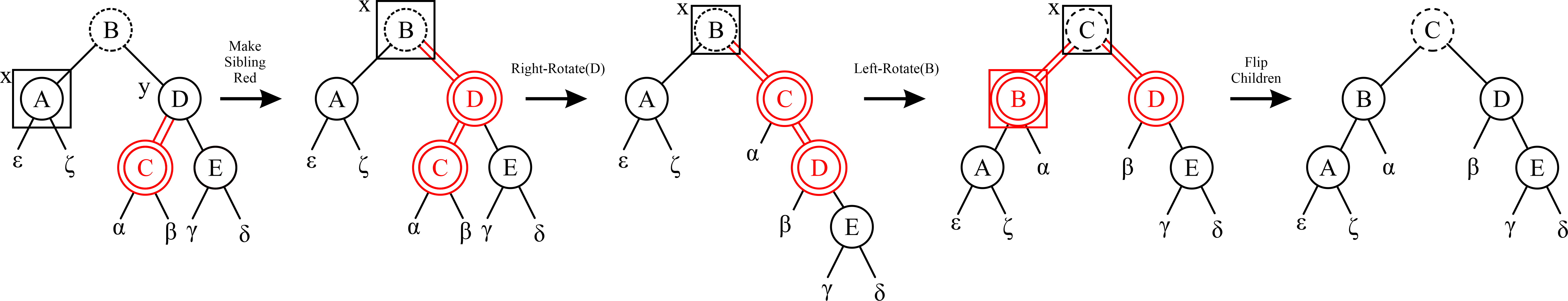}
\label{fig:2-3 RB-delete-case-2-2}
}
\vspace{0.5cm}
\subfloat[Rule (d) of classical delete algorithm is handled by the parity-seeking algorithm using main rule (b) followed by fixing rules (b) and (c).]{
\includegraphics[scale=\s]
{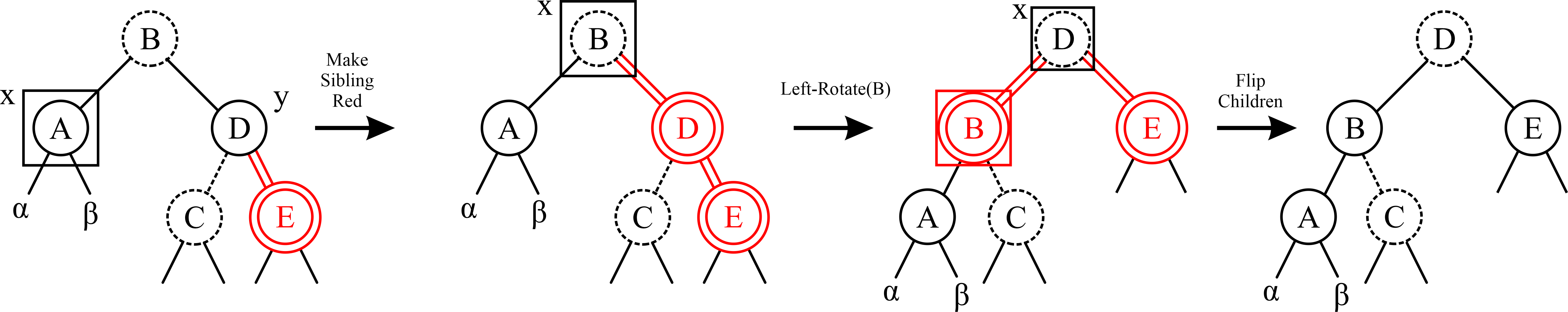}
\label{fig:2-3 RB-delete-case-2-3}
}\\
\vspace{0.5cm}
\subfloat[Rule (e) of classical delete algorithm is the same as the main rule (a) of the parity-seeking algorithm.]{
\includegraphics[scale=\sw, valign=c]{20191117_175736}
\label{fig:2-3 RB-delete-case-1}
}
\caption{The application of the parity-seeking and the classical delete algorithms \citep{cormen2009introduction} to 2-3-4 RB trees produces the same results, although following different rules. 
This shows that the proposed parity-seeking algorithm works for both 2-3 and 2-3-4 RB trees.}
\label{fig:2-3-4 RB-delete-black-leaf-equiv}
\end{figure}

In this section, we show that the proposed parity-seeking algorithm for 2-3 RB trees works without modification for 2-3-4 RB trees as well. 
To prove this, we show that the final result of applying the parity-seeking \algname{delete} algorithm and the classical \algname{delete} algorithm shown in Figure~\ref{fig:rbt-delete-case-3} are the same.
We consider the initial states of the rules of the classical delete algorithm and show that after several steps, both the classical and the parity-seeking delete algorithms reach the same state. 
Figure~\ref{fig:2-3-4 RB-delete-black-leaf-equiv} shows how the parity-seeking delete algorithm operates on each initial state of the classical delete algorithm. 
In all cases, except rule (c), the parity-seeking algorithm reaches the final state of the associated rule of the classical \algname{delete} algorithm. 
Rule (c) of classical delete algorithm prepares the state for a subsequent application of rule (d). 
For rule (c), the parity-seeking delete algorithm finally reaches a state that is the combination of the rules (c) and (d) of the classical delete algorithm. 
This shows that the parity-seeking delete algorithm works, without modification, for 2-3-4 RB trees as well.

\label{sec:experiments}

\section{Experiments}
\label{sec:experiments}
In this section, we experimentally compare our proposed 2-3 and 2-3-4 RB trees with classical RB trees and LLRB trees in inserting and deleting random sequences of numbers.
For LLRB trees, we started from the implementation of \citet{sedgewick2008left} in java\footnote{\url{https://algs4.cs.princeton.edu/33balanced/RedBlackBST.java.html}} and translated it to C++ for fair comparison. We were forced to modify the code slightly and handle some null references since the original java implementation crashed in our extensive tests. 
We implemented RB trees based on \citet{cormen2009introduction} with a \textit{nilSentinel} node, trying to make it similar to the elegantly concise implementation of LLRB. 
Then, we implemented our 2-3 and 2-3-4 RB trees with as few modifications as possible to the implementation of RB trees.
Our motivation for having a common basis for the implementation of RB, 2-3 RB, and 2-3-4 RB trees was to ensure that any difference in performance is solely due to algorithmic issues and all codes have been optimized to the same level.
For fair comparison, we added the \textit{nilSentinel} node to the implementation of LLRB, which helped in removing some conditional statements.
The implementations are available at \url{https://github.com/k-ghiasi/RedBlackTrees}.
All experiments have been performed on a UX310UQ notebook PC with an Intel(R) Core(TM) i7-6500U CPU @ 2.5GHz and 12 GB memory on a 64-bit windows 10 operating system. 
We report the number of rotations, visited nodes during the top-down pass, visited nodes  during the bottom-up pass, and the execution time.
For each size of data $n$, we randomly shuffled the integers $1$ to $n$ twice, and used one random shuffling for insertion and the other for search and deletion.
For size 1K we repeated the experiment 1000 times.
For sizes 10K and 100K we repeated the experiment 100 times. 
For sizes 1M and 10M we repeated the experiment 10 times. 

Table~\ref{table:exp-rotations+visits-insertion} shows the number of rotations, visited nodes during the top-down pass, and visited nodes during the bottom-up pass for the \algname{insert} algorithm of each variant of RB trees, normalized by $n \log n$ and multiplied by $1000$ for better readability. 
The most important observation is that the dominating factor is the number of visited nodes during the top-down pass. 
Since LLRB continues the bottom-up pass to the root, it has also an exceptionally high number of operations during the bottom-up pass. From this table, we can anticipate that the running time of LLRB should be almost twice that of other variants.
In addition, the average number of rotations in LLRB is almost 2 times of RB.
Comparing RB and 2-3 RB trees, we observe that the number of rotations in the \algname{insert} algorithm of 2-3 RB trees is almost $1.08$ times that of RB trees. 
Considering the slight difference in the number of rotations, and very similar number of visits during the top-down pass, we anticipate that the running time of the \algname{insert} algorithm for 2-3 RB and classical RB trees should be very close.

Table~\ref{table:exp-rotations+visits-deletion} shows the number of rotations, visited nodes during the top-down pass, and visited nodes during the bottom-up pass for the \algname{delete} algorithm of each variant of RB trees, normalized by $n \log n$ and multiplied by $1000$ for better readability. 
Again, the dominating factor is the number of visited nodes during the top-down pass. 
Since LLRB continues the bottom-up pass to the root, it has also an exceptionally high number of operations during the bottom-up pass. From this table, we can anticipate that the running time of LLRB should be almost triple times that of other variants.
In addition, the average number of rotations in LLRB is almost $20$ times of RB.
Comparing RB and 2-3 RB trees, we observe that the number of rotations in the \algname{delete} algorithm of 2-3 RB trees is almost $1.04$ times of that of RB trees. 
Considering the slight difference in the number of rotations, and very similar number of visits during the top-down pass, we anticipate that the running time of the \algname{delete} algorithm for 2-3 RB and classical RB trees should be very close.
In addition, we see that the number of visited nodes during the top-down pass is slightly lower for 2-3 RB trees compared to RB trees. 
This shows that the average height in 2-3 RB trees is slightly lower than classical RB trees \footnote{To see how this happens in an example, compare page 66 of 
\url{https://profsite.um.ac.ir/~k.ghiasi/publications/RBT2022/2-3-psrbt-insert-images.pdf}
and page 62 of
\url{https://profsite.um.ac.ir/~k.ghiasi/publications/RBT2022/clrs-insert-images.pdf}
}.

Table~\ref{table:exp-time-insert-delete} reports the average running time for the \algname{insert} and \algname{delete} algorithms for RB, LLRB, and 2-3 RB, and 2-3-4 RB trees, normalized by $n \log n$. 
As we anticipated, the running time of RB, 2-3 RB, and 2-3-4 RB trees are almost equal for both \algname{insert} and \algname{delete} algorithms.
while the running time of LLRB trees is almost twice of them.
This shows that the number of rotations is not an appropriate unit for measuring the running time of red-black trees as it does not reflect the actual running time. Although, our motivation for introducing the parity-seeking \algname{delete} algorithm was pedagogical, we observe that the resulting algorithm is also very efficient.

Table~\ref{table:exp-time-visits-search} reports the average running time and the number of visited nodes during the top-down pass when searching for a key at random. As can be seen, RB, 2-3 RB, and 2-3-4 RB variants perform almost identically on the search task. However, a t-test statistical analysis shows that the performance of LLRB is significantly worse than other variants.

\begin{table} [H]
\caption{Average number of rotations, top-down visits, and bottom-up visits of each variant of red-black trees during random insertion of different sizes.
  Results for 2-3-4 RB trees are not tabulated since they are identical to that of RB trees.
  Results are normalized by $n \log n$.
  For better visibility, the values are multiplied by $1000$.
  The column $n$ shows the number of inserted random integers. 
}
\begin{center}
\small
\begin{tabular}{l||c|c|c||c|c|c||c|c|c}\hline
& \multicolumn{3}{c||} {\small{Rotations} } 
 &     \multicolumn{3}{c||} {\small{Top-down visits}}
 &     \multicolumn{3}{c} {\small{Bottom-up visits}}
 \\\hline
n	&RB&	LLRB&	2-3 RB&	RB&	LLRB&	2-3 RB&	RB&	LLRB&	2-3 RB\\\hline
1K &	 $193$ &	 $569$ &	 $208$ &	 $3257$ &	 $3283$ &	 $3251$ &	 $298$ &	 $2950$ &	 $718$ 	 \\
10K &	 $146$ &	 $430$ &	 $157$ &	 $3289$ &	 $3323$ &	 $3283$ &	 $225$ &	 $3073$ &	 $541$ 	 \\
100K &	 $117$ &	 $345$ &	 $126$ &	 $3308$ &	 $3350$ &	 $3302$ &	 $180$ &	 $3150$ &	 $433$ 	 \\
1M &	 $97$ &	 $287$ &	 $105$ &	 $3320$ &	 $3364$ &	 $3315$ &	 $150$ &	 $3197$ &	 $361$ 	 \\
10M &	 $83$ &	 $246$ &	 $90$ &	 $3330$ &	 $3379$ &	 $3325$ &	 $129$ &	 $3236$ &	 $309$ 	 \\
   \hline
\end{tabular}
\end{center}
\label{table:exp-rotations+visits-insertion}
\end{table}

\begin{table} [H]
\caption{Average number of rotations, top-down visits, and bottom-up visits of each variant of red-black trees during random deletion from the trees of Table~\ref{table:exp-rotations+visits-insertion}. Deletion has been continued until the tree became empty.
  For 2-3-4 RB trees, the number of rotations and the number of top-down visits are not tabulated since they are identical to that of RB trees.
  Results are normalized by $n \log n$.
  For better visibility, the values are multiplied by $1000$.
  The column $n$ shows the initial size of the red-black trees. 
}
\begin{center}
\small
\begin{tabular}{l||c|c|c||c|c|c||c|c|c|c}\hline
& \multicolumn{3}{c||} {\small{Rotations} } 
 &     \multicolumn{3}{c||} {\small{Top-down visits}}
 &     \multicolumn{4}{c} {\small{Bottom-up visits}}
 \\\hline
n	&RB&	LLRB&	2-3 RB&	RB&	LLRB&	2-3 RB&	RB&	LLRB&	2-3 RB & 2-3-4 RB\\\hline
1K &	 $126$ &	 $2540$ &	 $131$ &	 $2899$ &	 $3898$ &	 $2890$ &	 $240$ &	 $3245$ &	 $339$ &	 $259$ 	 \\
10K &	 $95$ &	 $2940$ &	 $99$ &	 $3018$ &	 $3977$ &	 $3009$ &	 $181$ &	 $3478$ &	 $256$ &	 $196$ 	 \\
100K &	 $76$ &	 $3177$ &	 $79$ &	 $3089$ &	 $4017$ &	 $3081$ &	 $145$ &	 $3617$ &	 $205$ &	 $157$ 	 \\
1M &	 $63$ &	 $3315$ &	 $66$ &	 $3137$ &	 $4046$ &	 $3130$ &	 $121$ &	 $3713$ &	 $171$ &	 $131$ 	 \\
10M &	 $54$ &	 $3377$ &	 $56$ &	 $3173$ &	 $4079$ &	 $3166$ &	 $104$ &	 $3793$ &	 $147$ &	 $112$ 	 \\
   \hline
\end{tabular}
\end{center}
\label{table:exp-rotations+visits-deletion}
\end{table}

\begin{table} [H]
\caption{Average running time (in nanoseconds) for the insertion and the deletion algorithms of each variant of red-black trees, normalized by $n \log n$. 
The column $n$ shows the number of random integers inserted and removed from the red-black trees. }
\begin{center}
\small
\begin{tabular}{l||c|c|c|c||c|c|c|c}\hline
 & \multicolumn{4}{c||} {\small{Normalized Average Insertion Time}} & 
       \multicolumn{4}{c} {\small{Normalized Average Deletion Time}} \\\hline
n & RB&	LLRB&	2-3 RB&	2-3-4 RB&	RB&	LLRB& 2-3 RB&	2-3-4 RB\\\hline
1K &	 $52$ &	 $67$ &	 $52$ &	 $53$ &	 $41$ &	 $93$ &	 $41$ &	 $41$ \\ 
10K &	 $51$ &	 $70$ &	 $52$ &	 $52$ &	 $43$ &	 $105$ &	 $44$ &	 $42$ \\ 
100K &	 $58$ &	 $80$ &	 $59$ &	 $61$ &	 $52$ &	 $123$ &	 $53$ &	 $54$ \\ 
1M &	 $105$ &	 $148$ &	 $107$ &	 $108$ &	 $112$ &	 $219$ &	 $117$ &	 $116$ \\ 
10M &	 $161$ &	 $213$ &	 $165$ &	 $162$ &	 $183$ &	 $319$ &	 $188$ &	 $185$ \\  
   \hline
\end{tabular}
\end{center}
\label{table:exp-time-insert-delete}
\end{table}

\begin{table} [H]
\caption{Average running time (in nanoseconds) and the number of top-down visits in the course of searching a random element for each variant of red-black trees, normalized by $n \log n$. 
The column $n$ shows the number of items in the sought red-black trees. }
\begin{center}
\small
\begin{tabular}{l||c|c|c|c||c|c|c|c}\hline
 & \multicolumn{4}{c||} {\small{Normalized Average Search Time}} & 
       \multicolumn{4}{c} {\small{Normalized Average \# top-down visits}} \\\hline
n & RB&	LLRB&	2-3 RB&	2-3-4 RB&	RB&	LLRB& 2-3 RB&	2-3-4 RB\\\hline
1K &	 $21\pm2$ &	 $21\pm3$ &	 $21\pm2$ &	 $21\pm1$ &	 $3.08\pm0.01$ &	 $3.13\pm0.03$ &	 $3.08\pm0.01$ &	 $3.08\pm0.01$ \\ 
10K &	 $25\pm2$ &	 $25\pm5$ &	 $25\pm2$ &	 $25\pm2$ &	 $3.16\pm0.01$ &	 $3.20\pm0.02$ &	 $3.15\pm0.01$ &	 $3.16\pm0.01$ \\ 
100K &	 $37\pm2$ &	 $48\pm17$ &	 $37\pm1$ &	 $39\pm6$ &	 $3.20\pm0.01$ &	 $3.24\pm0.01$ &	 $3.20\pm0.01$ &	 $3.20\pm0.01$ \\ 
1M &	 $90\pm1$ &	 $98\pm5$ &	 $90\pm1$ &	 $92\pm5$ &	 $3.23\pm0.01$ &	 $3.28\pm0.02$ &	 $3.23\pm0.00$ &	 $3.23\pm0.01$ \\ 
10M &	 $146\pm1$ &	 $167\pm7$ &	 $147\pm1$ &	 $147\pm0$ &	 $3.25\pm0.00$ &	 $3.31\pm0.01$ &	 $3.25\pm0.00$ &	 $3.25\pm0.00$ \\  
   \hline
\end{tabular}
\end{center}
\label{table:exp-time-visits-search}
\end{table}

\section{Conclusions}
\label{sec:conclusions}
In this paper, we revisited 2-3 RB trees and introduced the parity-seeking \algname{delete} algorithm. Our goal was to introduce a pedagogically sound and easily understandable algorithm for deletion in red-black trees.
The proposed parity-seeking \algname{delete} algorithm is very natural and easily understandable. Specifically, the rationale behind the parity-seeking \algname{delete} algorithm is to balance the deficient subtree and its sibling by either fixing the deficient subtree or making the sibling also deficient, elevating the deficiency one level higher. 
In our experiments, we found that the performance of 2-3 RB trees is very close to classical RB trees both in the \algname{insert} and \algname{delete} operations. 
Besides, the introduced parity-seeking \algname{delete} algorithm also works for 2-3-4 RB trees and its performance is almost identical to the classic \algname{delete} algorithm of RB trees.
The goal of devising a simple yet efficient algorithm for the delete operation in red-black trees is finally achieved.

\section*{Author Contributions}
The parity-seeking \algname{delete} algorithm came to the mind of Kamaledin Ghiasi-Shirazi when he taught LLRB trees in his data structure course. He invited his former students, Taraneh Ghandi, Ali Taghizadeh, and Ali Rahimi-Baigi, to participate in the preparation of this paper. All authors validated the idea in common sessions, and Ali Taghizadeh, Ali Rahimi-Baigi, and Taraneh Ghandi implemented 2-3 RB along with the competing methods of RB and LLRB. 
Ali Taghizadeh and Ali Rahimi-Baigi carefully studied RB and LLRB trees and explained it to other members of the team.
The paper was initially written on the blackboard of a classroom in Persian, with all authors participating and discussing. The paper was then translated to English by Taraneh Ghandi and Kamaledin Ghiasi-Shirazi. All graphics have been produced by Taraneh Ghandi. Considering the extreme importance of the topic, Kamaledin Ghiasi-Shirazi re-implemented RB, 2-3 RB, and 2-3-4 RB trees in a unified framework for a fair comparison.
Kamaledin Ghiasi-Shirazi revised the manuscript and prepared the final manuscript. 
All authors carefully read and commented on the final manuscript.


\end{document}